\documentclass{aa} 
\RequirePackage{aas_macros}
\usepackage{graphicx}
\usepackage{natbib}
\bibpunct{(}{)}{;}{a}{}{,} 
\usepackage{txfonts}
\usepackage{aalongtable}
\usepackage{multirow}
\usepackage{placeins}
\usepackage{hyperref} 

\begin{document}
\def\r500{$r_{\mathrm{500}}$}
\def\P3P0{$P3/P0$}
\def\B_P3{$B_{P3}$}
\def\b_w{$B_{w}$}
\def\id_P3{$P3/P0_{ideal}$}
\def\i_w{$w_{ideal}$}
\def\C_B_P3{$B_{P3,c}$}
\def\c_b_w{$B_{w,c}$}
\def\p3_c{$P3/P0_{c}$}
\def\cp3{$P3/P0_c$}
\def\w_c{$w_{c}$}

   \title{Probing the evolution of the substructure frequency in galaxy clusters up to $z\sim1$}
   \author{A. Wei\ss mann\inst{1}, H. B\"{o}hringer\inst{1}, G. Chon\inst{1}}
   \institute{$^1$ Max-Planck-Institut f\"{u}r extraterrestrische Physik, Postfach 1312, Giessenbachstr., 85741 Garching, Germany\\ {\tt email: weissman@mpe.mpg.de}}
              
   \date{Received 18 March 2013 / Accepted 4 June 2013}

 \abstract
 {Galaxy clusters are the last and largest objects to form in the standard hierarchical structure formation scenario through merging of smaller systems. The substructure frequency in the past and present epoch provides excellent means for studying the underlying cosmological model.} 
 {Using X-ray observations, we study the substructure frequency as a function of redshift by quantifying and comparing the fraction of dynamically young clusters at different redshifts up to $z=1.08$. We are especially interested in possible biases due to the inconsistent data quality of the low-z and high-z samples.}
 {Two well-studied morphology estimators, power ratio \P3P0 and center shift $w$, were used to quantify the dynamical state of 129 galaxy clusters, taking into account the different observational depth and noise levels of the observations.}
 {Owing to the sensitivity of \P3P0 to Poisson noise, it is essential to use datasets with similar photon statistics when studying the \P3P0-z relation. We degraded the high-quality data of the low-redshift sample to the low data quality of the high-z observations and found a shallow positive slope that is, however, not significant, indicating a slightly larger fraction of dynamically young objects at higher redshift. The $w$-z relation shows no significant dependence on the data quality and gives a similar result.}
 {We find a similar trend for \P3P0 and $w$, namely a very mild increase of the disturbed cluster fraction with increasing redshifts. Within the significance limits, our findings are also consistent with no evolution.} 
 
\authorrunning{Wei\ss mann et al.}
\titlerunning{Probing the evolution of the substructure frequency in galaxy clusters up to $z\sim1$}
\maketitle
\titlerunning{short title}
\authorrunning{Names}
%

\section{Introduction}
The standard theory of structure formation predicts hierarchical growth from positive fluctuations in the primordial density field. Subgalactic scale objects decouple first, then collapse and virialize due to the greater amplitudes of the density fluctuations on small scales. They grow through merging, finally forming galaxy clusters, which are considered the largest virialized objects in the Universe. Galaxy cluster growth probes the evolution of density perturbations and directly traces the process of structure formation in the Universe. Galaxy clusters are thus important laboratories for studying and testing the underlying cosmological model \citep[e.g.][]{Borgani2008LN,Voit2005}. Especially important in this context is the study of the cluster mass function, whose evolution provides constraints on the linear growth rate of density perturbations. Using X-rays and analyzing the hot intracluster medium (ICM) that resides in the deep potential well of galaxy clusters, mass determination is based on the assumptions of hydrostatic equilibrium and spherical shape. These assumptions may be unsatisfactory for dynamically young objects showing multiple surface brightness peaks in the distribution of the ICM, however \citep[e.g.][]{Nelson2012, Rasia2012, Zhang2008}. In addition, the influence of dynamical activity such as merging on $L_\mathrm{X}$, $T_\mathrm{X}$ etc. needs to be known in detail to explain possible deviations from scaling relations for disturbed clusters \citep[e.g.][]{Pratt2009, Rowley2004, Chon2012} with the aim to reduce the errors in cosmological studies.

Observations of substructure and disturbed morphologies in the optical \citep[see e.g.][and references therein]{Girardi2002, West1990} and X-ray band \citep[for a review see e.g.][]{Buote2002} indicate that a large fraction of clusters is dynamically young and has not reached a relaxed state yet. It is therefore essential to quantify the fraction of disturbed clusters that reflects the formation rate and to probe higher redshifts to constrain cosmological parameters.

X-ray observations provide excellent probes for studying the dynamical state of clusters because the ICM traces their deep potential well.
Over the years, X-ray studies became very efficient in quantifying cluster structure, and a variety of X-ray morphology estimators was introduced \citep[for a review see][]{Rasia2012b}. However, only recently, larger samples of high-z observations of galaxy clusters became available and allowed statistical studies of the evolution of the substructure frequency up to $z\sim1$. Since then, several observational X-ray studies have shown a larger fraction of dynamically relaxed clusters at lower redshift than at $z>0.5$ \citep[e.g.][]{Mann2012,Andersson2009,Maughan2008, Hashimoto2007a, Bauer2005, Jeltema2005, Plionis2002, Melott2001}. A less clear evolution was found in hydrodynamical simulations, but higher merger rates at high redshift support the observational results \citep[e.g.][]{Burns2008, Jeltema2008,Kay2007,Rahman2006, Cohn2005}. 

Opening the window toward higher-redshift clusters is accompanied by the problem of the insufficient data quality of X-ray images in terms of net photon counts and background contribution. Exploring a broad redshift range directly translates into probing data with quite substantial quality differences. It is therefore not only essential to use well-studied morphology estimators but also to understand possible biases caused by uneven data quality.

In this work, we used two common X-ray substructure estimators, power ratio \P3P0 \citep{Buote1995} and center shift $w$ \citep{Mohr1993}, to study the relation between cluster structure and redshift up to $z=1.08$. To do so, we took advantage of the detailed study of the influence of net photon counts and background on the computation of \P3P0 and $w$ in our recently published work \citep{Weissmann2013}.   
\citet{Jeltema2005} presented the first analysis of the \P3P0-z relation using 40 X-ray selected luminous clusters in the redshift range $0.1<z<0.89$. Using different statistical measures, they reported on average higher \P3P0 for clusters with $z>0.5$ than for low-z objects. While they accounted for the bias caused by photon noise and background, they did not fully consider the strong decrease of data quality at higher redshifts and overestimated the \P3P0-z relation. In addition to using a larger sample, we explored possible biases caused by different observational depths in the low-z and high-z samples and determined how to account for them when analyzing the \P3P0-z and $w$-z relation.

The paper is organized as follows. We characterize the sample and briefly discuss the data reduction process in Sect. \ref{Sect2}. In Sect. \ref{Sect3} we introduce the morphology estimators \P3P0 and $w$ used in this work. Sect. \ref{DataQuality} summarizes how we degraded the high-quality data of the low-z sample to match the high-z observations. We give results in Sect. \ref{Results}, including a detailed study of the influence of the different data quality in samples. Previous studies and the effect of cool cores are discussed  in Sect. \ref{Discussion}. We finally conclude with Sect. \ref{Conclusion}. Throughout the paper, the standard $\Lambda$CDM cosmology was assumed: $H_{0}$=70 km s$^{-1}$ Mpc$^{-1}$, $\Omega_{\Lambda}$=0.7, $\Omega_{\mathrm{M}}$=0.3.

\section{Observations and data reduction}
\label{Sect2}
\begin{figure}
 \begin{center}
  \includegraphics[width=\columnwidth]{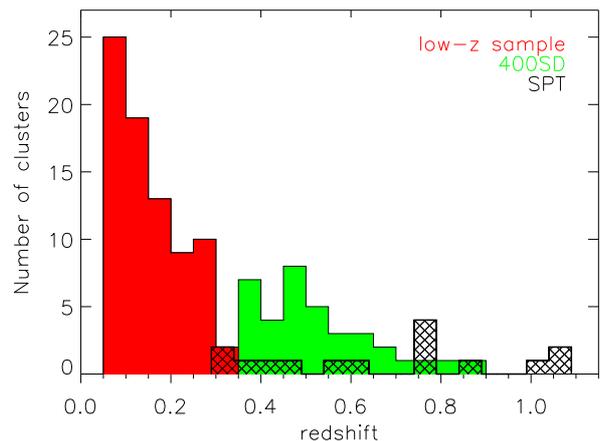}
 \end{center}
\caption{Redshift distribution of the low-z (red filled histogram), high-z 400SD (green filled histogram), and high-z SPT sample (black dashed histogram) .}
\label{zoverview}
\end{figure}

\begin{figure}
 \begin{center}
  \includegraphics[width=0.49\columnwidth]{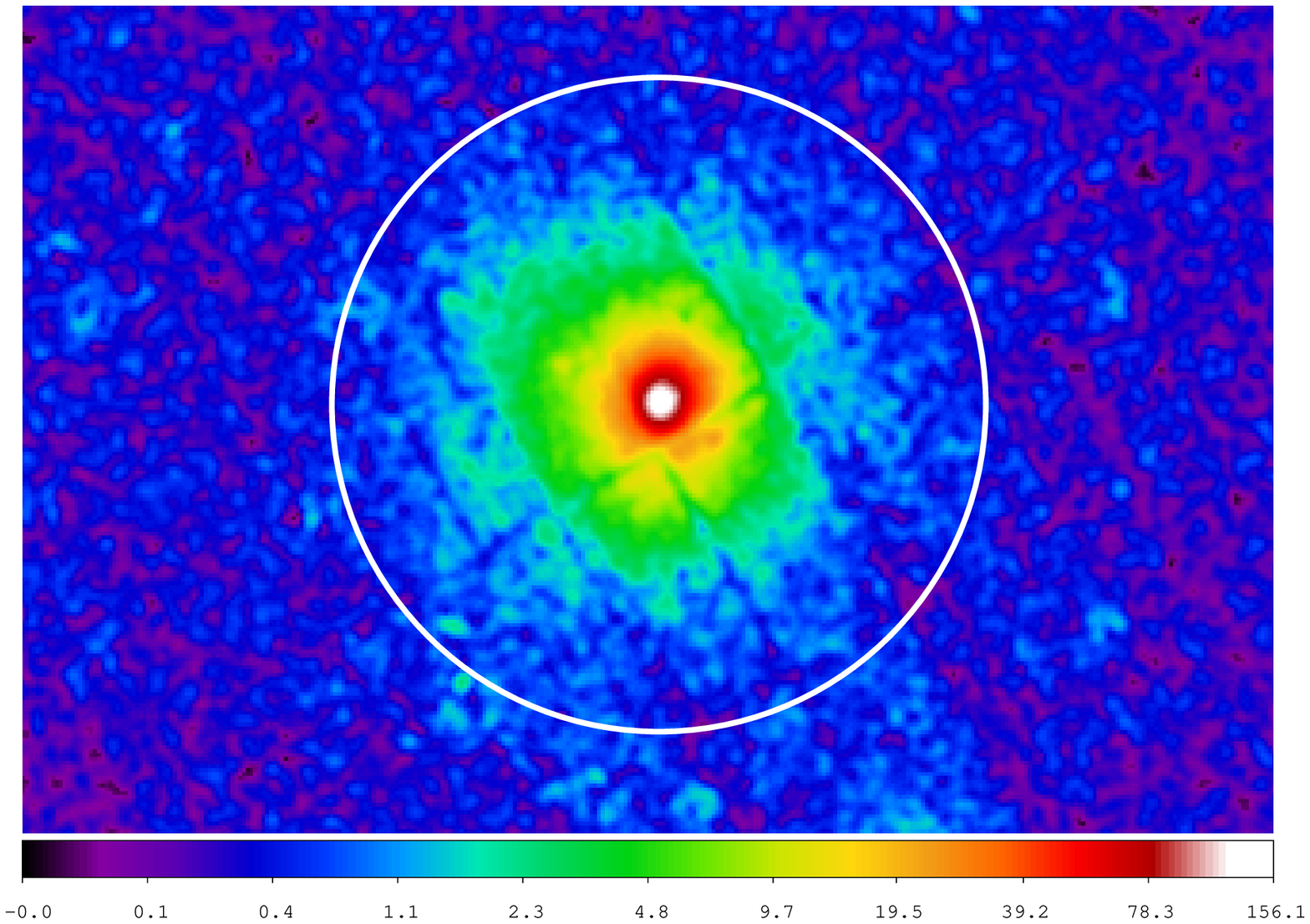}
  \includegraphics[width=0.49\columnwidth]{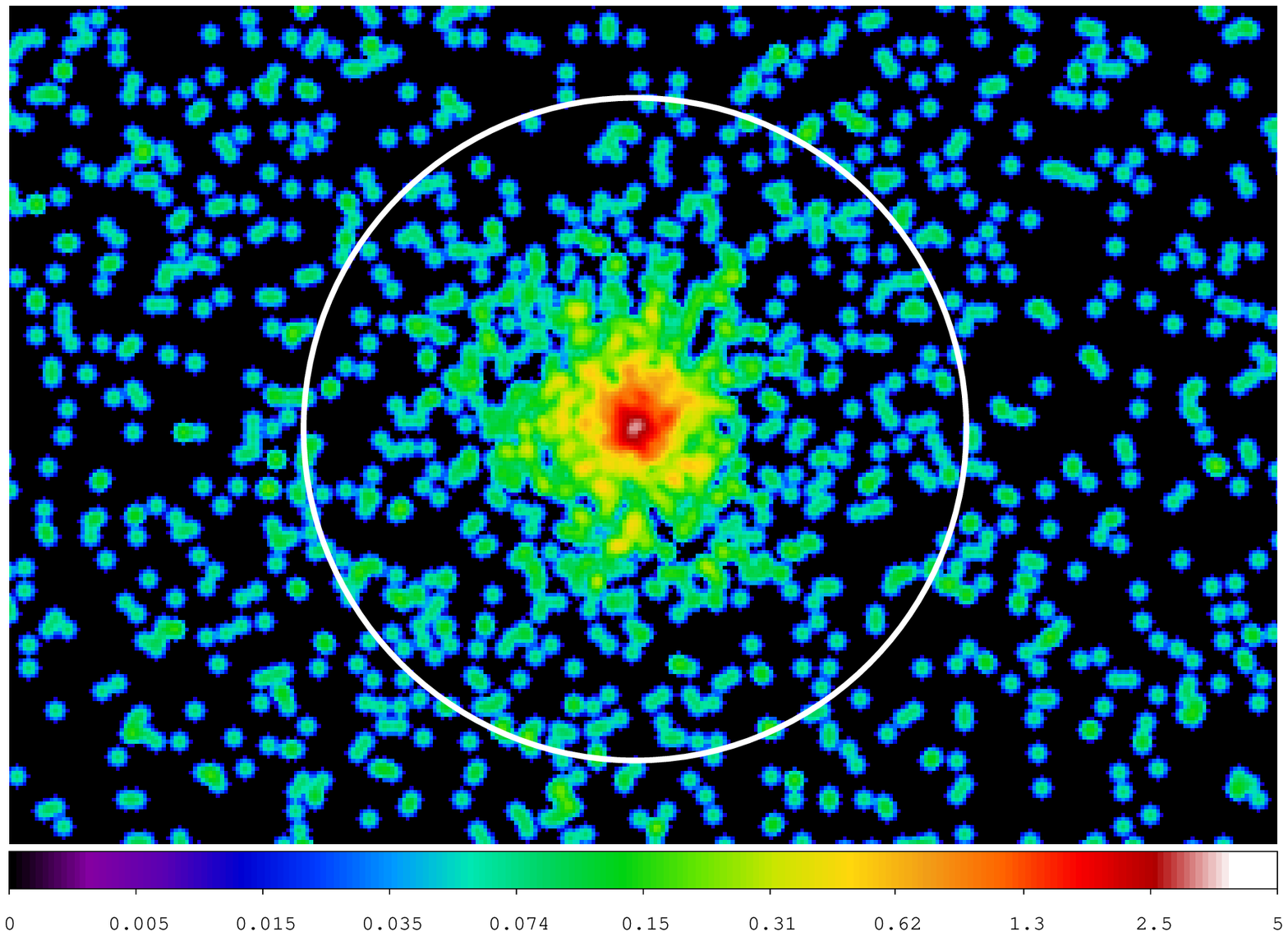}
  \includegraphics[width=0.49\columnwidth]{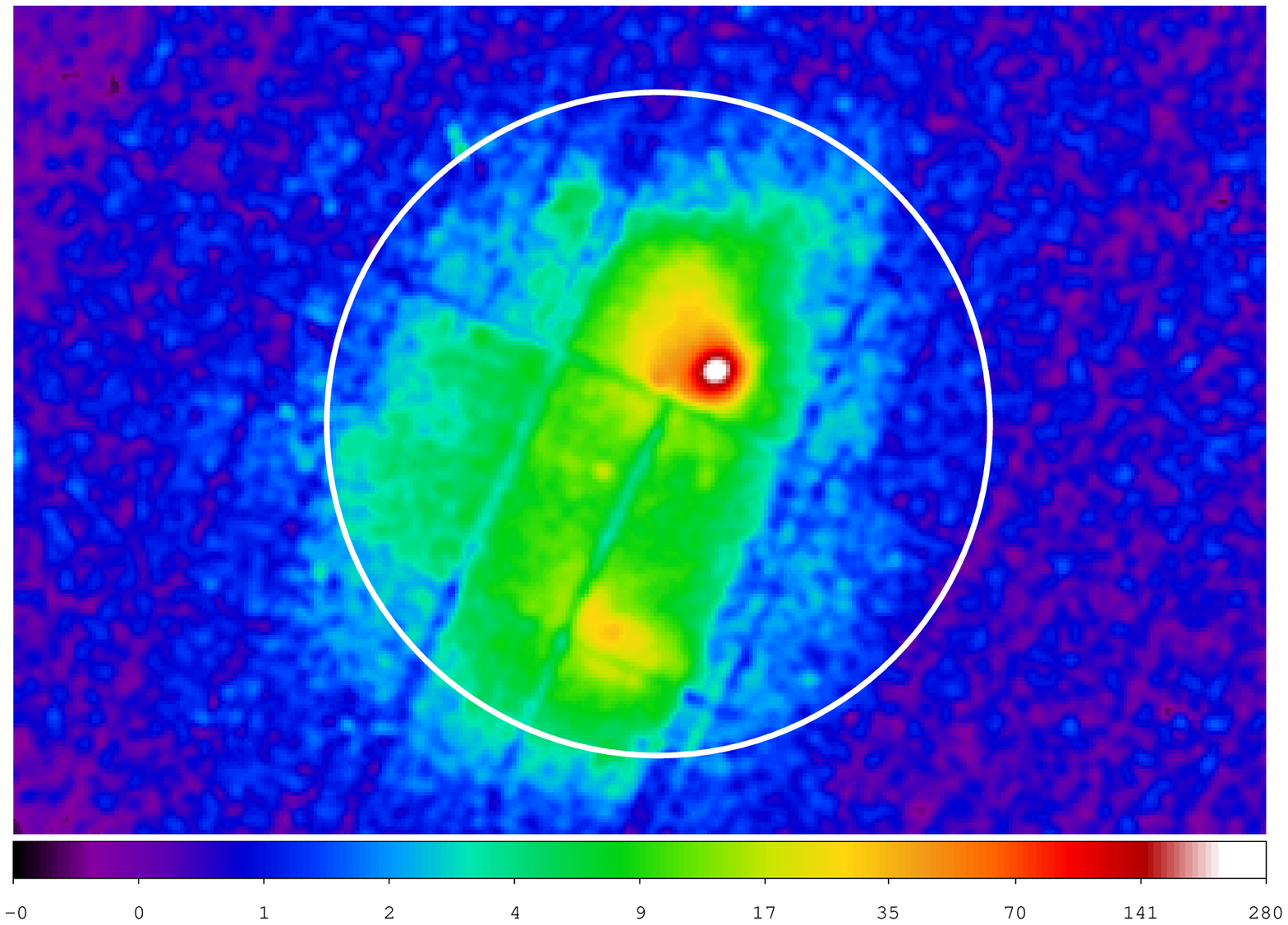}
  \includegraphics[width=0.49\columnwidth]{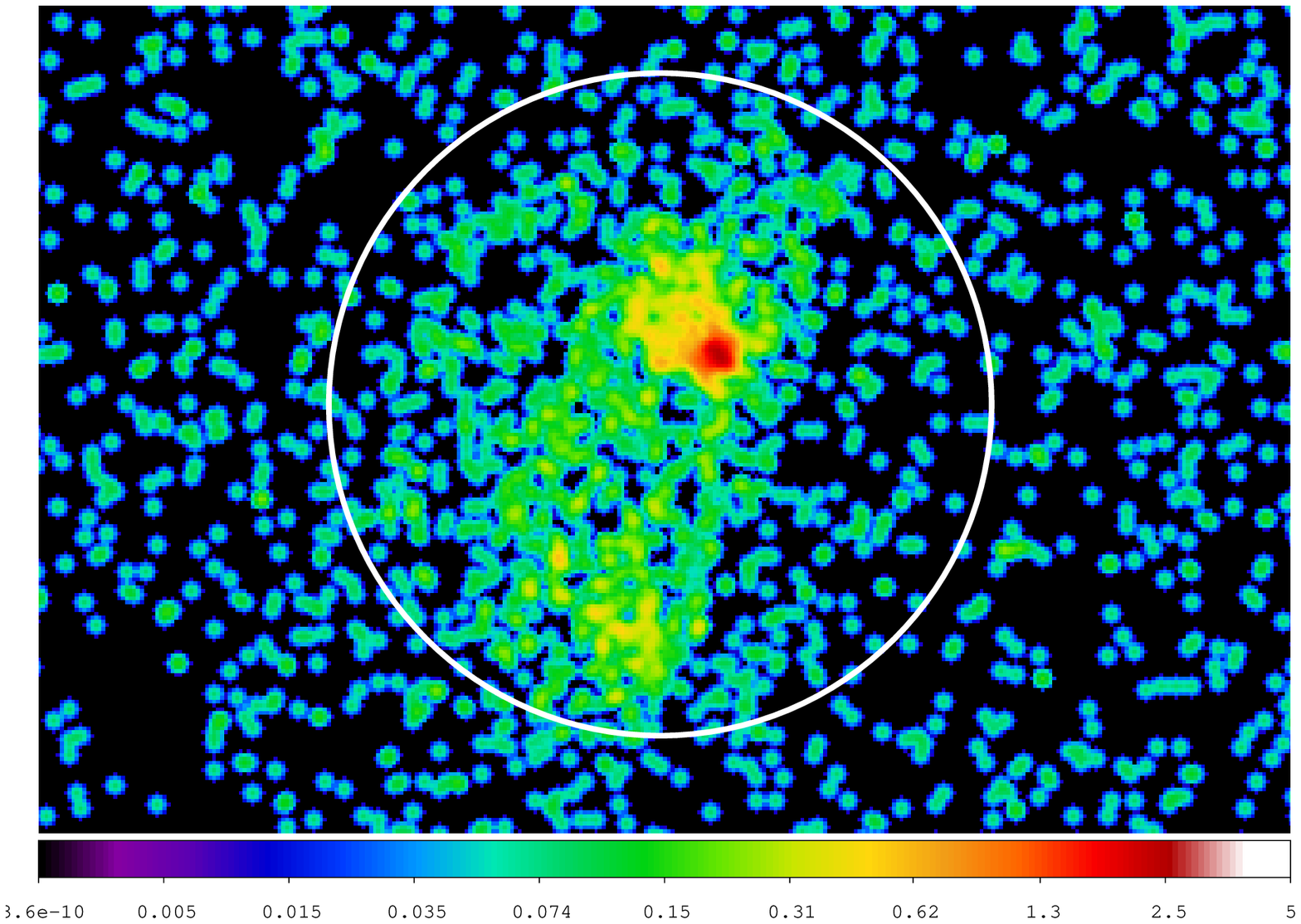}
\end{center}
\caption{Examples of the background-included, point-source-corrected smoothed X-ray images of the low-z sample. Left: high-quality (undegraded) images. Right: degraded images (for details see Sect. \ref{degrading}). Top panels: A963 - relaxed galaxy cluster at $z=0.21$ with \P3P0=$(1.77\pm1.42)\times10^{-8}$ and $w=(4.40\pm0.30)\times10^{-3}$, non-significant detection after degrading. Bottom panels: A115 - merging cluster at $z=0.20$ with \P3P0=$(5.33\pm0.19)\times10^{-6}$ and $w=(8.54\pm0.05)\times10^{-2}$, significant detection after degrading. The circle indicates \r500.}                                                                                                                                               
\label{lowzim}
\end{figure}

In this section we discuss the three samples used for our study: the low-z sample and the high-z subsamples of the 400SD and SPT surveys. An overview of the redshift distribution is shown in Fig. \ref{zoverview}. Table \ref{statistics} summarizes the sample statistics including the number of clusters, the redshift range, the mean net photon counts within \r500, and the mean net- (signal-)to-background photon counts ratio S/B. This table is discussed in more detail in Sect. \ref{DataQuality}, where we concentrate on the problem of the data quality. Details of the galaxy clusters and observational properties are given in Table \ref{SampleInfo}. \r500 was calculated for all clusters using the formula given by \citet{Arnaud2005}. The temperature and redshift values were taken from previous works as indicated in Table \ref{SampleInfo}. For a full gallery of the X-ray images of the galaxy clusters used in this study we refer to \citet{Weissmann2013} for the low-z sample, the website of the 400d$^2$ cluster survey\footnote{http://hea.iki.rssi.ru/400d/catalog/} for the high-z 400SD objects, and to \citet{Andersson2011} for the high-z SPT clusters. To give an impression of the substructure values and the data quality, we provide a few examples of background-included, point-source-corrected smoothed X-ray images in Fig. \ref{lowzim} (left panels) for the low-z sample and in Fig. \ref{highzim} for the high-z samples.

\subsection{Low-z cluster sample}

The low-redshift sample (short: low-z) was previously used and discussed in detail in \citet[][W13 hereafter]{Weissmann2013}. For our current work, we excluded two clusters that were part of the W13 sample: RXJ1347-1145 and RXCJ0516-5430. RXJ1347-1145 was omitted because of its high redshift of $z=0.45$ and because we did not want to add this cluster to the high-z samples with defined origin. RXCJ0516-5430 or SPT-CLJ0516-5430 ($z=0.29$) was already part of the high-z SPT sample. We thus excluded it from the low-z sample because of its high redshift. \\
The low-z sample now comprises 78 archival XMM-Newton observations of galaxy clusters covering redshifts between 0.05 and 0.31, with $\langle z \rangle=0.15$. The clusters were drawn from several well-known samples observed with XMM-\emph{Newton} (for details see Table \ref{SampleInfo}): REXCESS \citep{Boehringer2007}, LoCuSS \citep[Smith et al., ][]{Zhang2008}, the Snowden Catalog \citep{Snowden2008}, the REFLEX-DXL sample \citep{Zhang2006}, and \citet{Buote1996}. The clusters were chosen to be well-studied, nearby ($0.05<z<0.31$), and publicly available (in 2009) in the XMM-Newton science archive\footnote{http://xmm.esac.esa.int/xsa/}. In addition, we required \r500 to fit on the detector. The calculation of \r500 using the formula of \citet{Arnaud2005} led to slightly different \r500 and hence \P3P0 and $w$ values to those quoted in W13. The differences are small, however. This merged low-z sample has no unique selection function, but a wide spread in luminosity, temperature, and mass. A large part of the clusters comes from representative samples such as REXCESS and LoCuSS and we therefore expect the sample to have a very roughly representative character. To check in more detail that no bias effect is introduced by the merged sample, we also performed all tests with the 31 REXCESS clusters only. The results are consistent with the full low-z sample and we therefore do not quote them in detail.

\subsection{High-z cluster samples}

In the high-redshift range, we used two samples to account for possible selection effects and performed our analysis on each sample individually: the X-ray-selected high-z subsample from the 400SD survey \citep[][]{Burenin2007,Vikhlinin2009} and the SZ-selected subsample from SPT discussed in \citet{Andersson2011}. \\

The high-z 400SD sample (short: 400SD) forms a complete subsample of the $z>0.35$ clusters from the 400SD survey. It is composed of 36 objects in the $0.35<z<0.89$ range and was selected as a quasi-mass-limited sample at $z>0.5$. This was done by requiring a luminosity above a threshold of $L_\mathrm{X,min}=4.8\times10^{43} (1+z)^{1.8}$ erg s$^{-1}$. All 36 400SD clusters were observed with CHANDRA and are publically available in the CHANDRA archive\footnote{http://cxc.cfa.harvard.edu/cda/}. Several authors \citep[e.g.][]{Santos2010a} have raised the question whether there might be a possible bias in the 400SD sample due to the detection algorithm. This may result in a lack of concentrated clusters compared with other high-redshift samples such as the Rosat Deep Cluster Survey \citep[RDCS,][]{Rosati1998} or the Wide Angle ROSAT Pointed Survey \citep[WARPS,][]{Jones1998}. We accounted for these effects by using the high-z SPT sample for comparison.\\

The high-z SPT sample (short: SPT) is a subsample of the first SZ-selected cluster catalog, obtained from observations of 178 deg$^{2}$ of sky surveyed by the South Pole Telescope (SPT). \citet{Vanderlinde2010} presented a significance-limited catalog of 21 SZ-detected galaxy clusters of which 15 objects with SZ-detection-significance above 5.4 were selected for an X-ray follow-up program. This subsample covers the redshift range $0.29<z<1.08$. The majority of the clusters was observed with CHANDRA, but for three objects we used XMM-Newton data because no CHANDRA data are available (SPT-CLJ2332-5358 and SPT-CLJ0559-5249) or because of the better photon statistics of the XMM-Newton observation (SPT-CLJ0516-5430). This results in 12 CHANDRA and 3 XMM-Newton observations (for details see Table \ref{SampleInfo}, Column 9).

\begin{figure}
 \begin{center}
  \includegraphics[width=0.49\columnwidth]{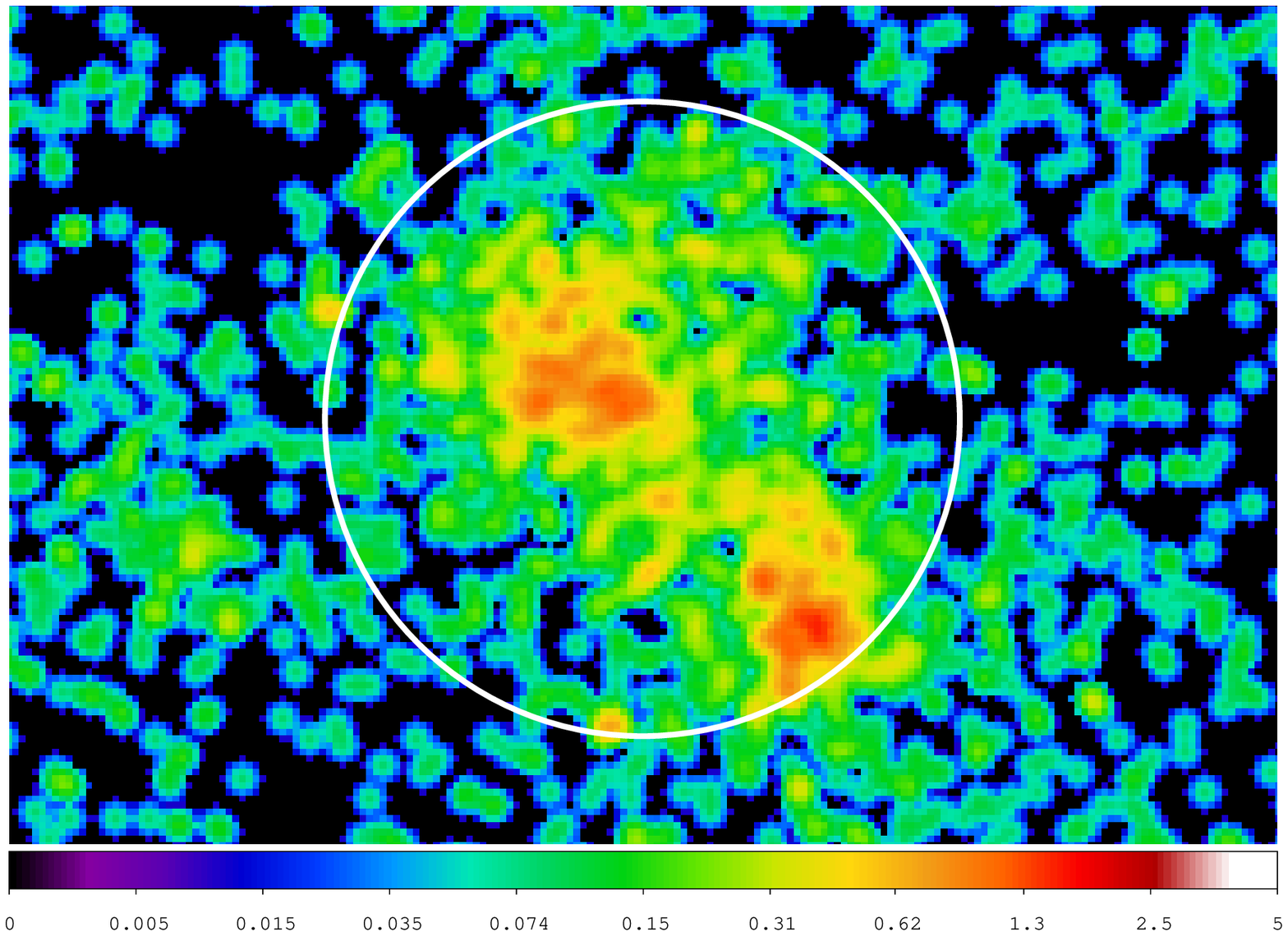}
  \includegraphics[width=0.49\columnwidth]{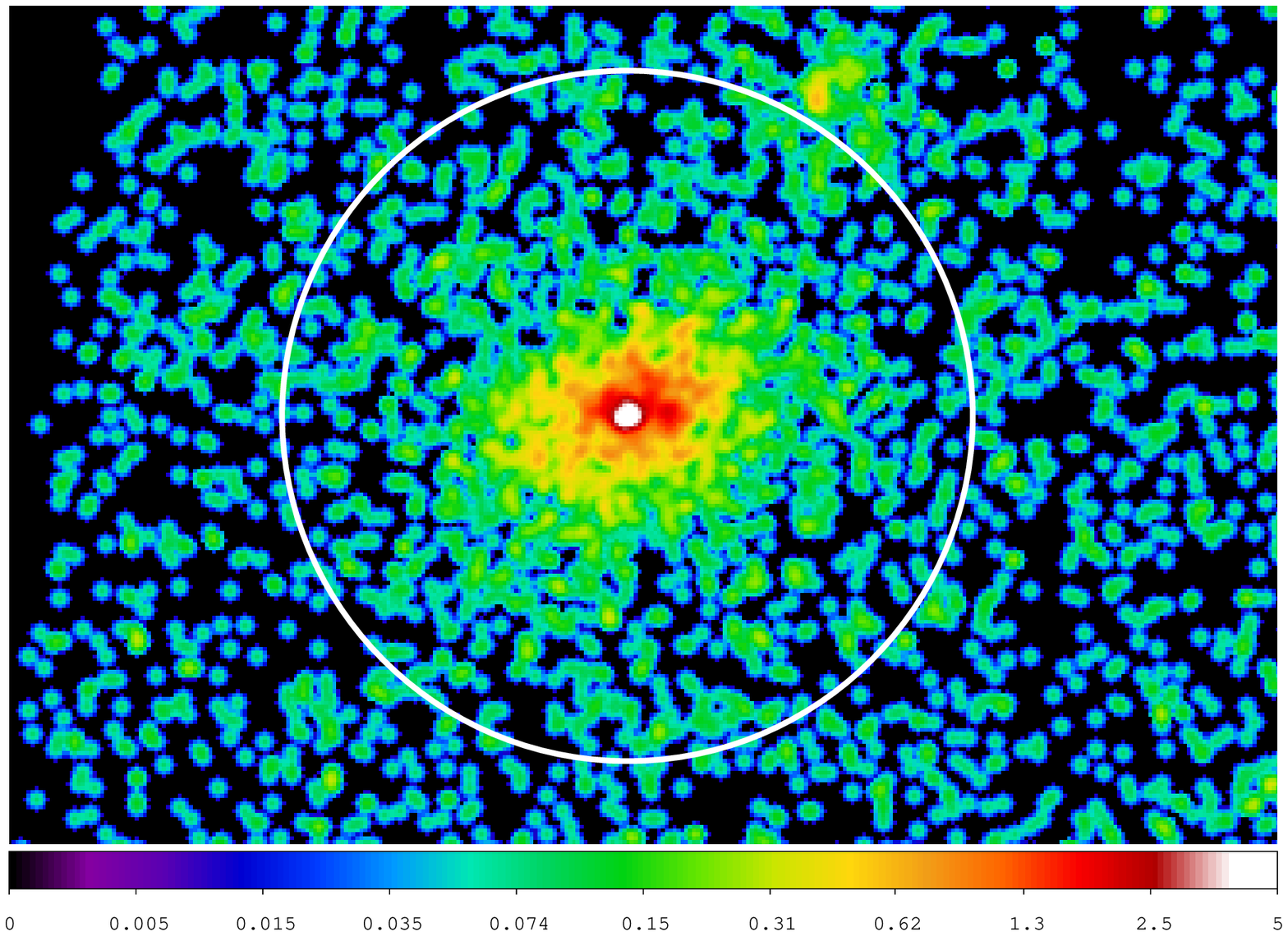}
\end{center}
\caption{Examples of the background-included, point-source-corrected smoothed X-ray images of the high-z samples. Left: 0152-1358 - very structured cluster at $z=0.83$ with \P3P0=$(5.76\pm0.95)\times10^{-5}$ and $w=(6.64\pm0.57)\times10^{-2}$. This 400SD cluster has the highest \P3P0 value and is marked by a circle in Figs. \ref{P3_orig}-\ref{original_data_w}. Right: SPT-CLJ0509-5342 - rather relaxed SPT cluster at $z=0.46$ with a non-significant detection in \P3P0 and $w=(3.13\pm1.33)\times10^{-3}$. The circle indicates \r500.}                                                                                                                                               
\label{highzim}
\end{figure}

\begin{table}
\begin{center}
\begin{tabular}{|l|c|c|c|}
\hline& & &\\[-1.5ex]
\multicolumn{1}{|c|}{} & \multicolumn{1}{|c|}{Low-z} &\multicolumn{1}{|c|}{400SD}&\multicolumn{1}{|c|}{SPT}\\ [0.5ex]
\hline & & &\\[-1.5ex] 
Number of clusters & 78 & 36 & 15\\
Redshift range & 0.05-0.31 & 0.35-0.89 & 0.29-1.08 \\
Mean net photon counts & 96997 & 1203 & 1735 \\
Mean S/B & 6.6 & 3.7 & 3.2 \\
\hline& & &\\[-1.5ex]
\P3P0 $> 0$ & 68 & 22 & 11 \\ 
Upper limits & 14 & 21 & 8 \\
$S_{P3} > 1$ & 55 & 6 & 3 \\ 
$S_{P3} > 3$ & 28 & 2 & 0 \\ 
Mean $S_{P3}$ & 2.6 & 0.5 & 0.4\\
Median $S_{P3}$ & 2.0 & 0.2 & 0.1\\
\hline& & &\\[-1.5ex]
$w > 0$ & 78 & 35 & 15 \\ 
Upper limits & 1 & 3 & 1 \\
$S_{w} > 1$ & 77 & 28 & 13 \\ 
$S_{w} > 3$ & 74 & 19 & 5 \\ 
Mean $S_{w}$ & 23.0 & 3.8 & 3.4\\
Median $S_{w}$ & 17.0 & 3.8 & 2.6\\
\hline
\end{tabular}
\\[1.5ex] 
\caption{Overview of the data quality of the samples. Mean net photon counts and mean S/B calculated within \r500 of the reduced and point source corrected X-ray image. S/B gives the ratio of net photon counts (signal) to background photon counts. \P3P0 and $w$ are computed in the \r500 aperture. $P3/P0>0$ and $w>0$ include all clusters with positive corrected substructure values, including positive non-significant detections. For clusters with non-significant results, we quote upper limits, which are taken as the sum of the non-significant result (or zero for a neg. corrected \P3P0 or $w$) and the 1-$\sigma$ error (for details see Sect. \ref{Sect3}). The significance $S$ is computed as the ratio of \P3P0 or $w$ with respect to its error. This table is discussed in more detail in Sect. \ref{DataQuality}.}
\label{statistics}
\end{center}
\end{table}

\subsection{Data reduction}

The 78 low-z and additional 3 high-z SPT XMM-\emph{Newton} observations (SPT-CLJ2332-5358, SPT-CLJ0559-5249 and SPT-CLJ0516-5430) were taken from the public XMM-Newton Science archive and were analyzed with the XMM-\emph{Newton} SAS\footnote{Science Analysis Software: http://xmm.esa.int/sas/} in the well-established standard 0.5-2 keV band, which covers most of the cluster signal. The low-z clusters and SPT-CLJ0516-5430 were reduced prior to this study using SAS v. 9.0.0, while we used v. 12.0.1 for the other two SPT objects. In both cases we followed the data reduction recipe described in detail in \citet{Boehringer2009,Boehringer2007}, except for the point source removal. Point sources were detected with the SAS task \textit{ewavelet} in the combined image from all three detectors to increase the sensitivity of the point source detection. However, we removed the point sources from each detector image in the 0.5-2 keV band individually and refilled the gaps using the CIAO\footnote{CHANDRA Interactive Analysis of Observations software package: http://cxc.harvard.edu/ciao/} task \textit{dmfilth}. In the next step we subtracted the background, which was obtained from a vignetting model fit to a source-excised, hard-band-scaled blank sky field from the point-source-corrected images and combined them. This method yields point-source-corrected images without visible artifacts of the cutting regions.\\

The high-z CHANDRA observations of the 400SD and SPT sample were treated as follows. A standard data reduction in the 0.5-2 keV band was performed using the CIAO software package v4.4 and CALDB v4.4.7. This band was chosen to match the XMM-\emph{Newton} data. For each observation, the level = 1 event file was reprocessed using \emph{chandra\_repro}, including amongst others the detection of afterglows, the generation of a new bad pixel file and corrections for differing gains across the CCDs, time-dependent gain, and charge transfer inefficiencies (CTIs). For observations taken in the VFAINT mode, we applied the additional background cleaning using the task \emph{acis\_process\_events} while setting \emph{check\_vf\_pha=yes}. This procedure uses the outer 5 x 5 pixel (instead of 3 x 3 for FAINT) event island to search for potential cosmic-ray background events. Flared periods were excluded from the level = 2 event file using \emph{lc\_clean}. We created images in the 0.5-2 keV range and used \emph{fluximage} to generate monochromatic 1 keV exposure maps. Point sources were detected and removed using \emph{dmfilth}, which also refills the excised regions. For the background, blank-sky event files were reprojected, scaled to the exposure time of the flare-cleaned observation, restricted to the 0.5-2 keV range and binned with a factor of 4 to match the observations. When there were several pointings per cluster, we reduced the observations individually, but detected point sources on the merged 0.5-2 keV image. Images and exposure maps were merged using \emph{reproject\_image}. 


\section{Morphological analysis}
\label{Sect3}

We used power ratios and center shifts as morphology estimators for our analysis. The power ratio method was introduced by \citet{Buote1995} to quantify the amount of substructure in a cluster and its dynamical state. The powers are based on a 2D multipole expansion of the cluster's gravitational potential and are evaluated within a certain aperture radius (e.g. \r500). It is already well established that the normalized hexapole of the X-ray surface brightness, \P3P0, is sensitive to asymmetries on scales of the aperture radius and provides a useful measure of the dynamical state of a cluster \citep[e.g. ][W13]{Jeltema2005, Buote1995, Boehringer2009, Chon2012}. 
Moreover, the center shift parameter $w$ \citep[e.g. ][W13]{Ohara2006, Mohr1993, Boehringer2009, Chon2012} characterizes the morphology of the cluster X-ray surface brightness. It measures the shift of the centroid, defined as the center of mass of the X-ray surface brightness, with respect to the X-ray peak in different apertures. The X-ray peak was determined from an image smoothed with a Gaussian with $\sigma$ of 8 arcseconds. The offset of the X-ray peak from the centroid was then calculated for ten aperture sizes (0.1-1 \r500) and the final parameter $w$ obtained as the standard deviation of the different center shifts in units of \r500. Unless stated otherwise, all presented \P3P0 and $w$ values were calculated within an aperture of \r500 and including the central region. However, we exclude the central 0.1 \r500 region when we calculated the X-ray centroid for the discussion in Sect. \ref{effectcoolcores} to study possible effects of cool cores.\\

Both morphology estimators were discussed in our previous paper W13, where we studied the influence of background and shot noise on \P3P0 and $w$ as a function of photon counts and presented a method to correct for these effects. In short, we first subtract the moments of the background image from those of the full (background-included) image to obtain a background-corrected power ratio. In a second step, we correct the bias caused by shot noise using repoissonized realizations of the cluster image. For $w$ we subtract the background pixel values before calculating the position of the X-ray peak and centroid and estimate the shot noise bias analogous to the power ratios. For very regular clusters or observations highly influenced by noise, we sometimes overestimate the bias and obtain negative corrected \P3P0 and $w$ values with errors exceeding the negative value. We call such results non-significant detections. Substructure values that are positive after the bias correction, but have a 1-$\sigma$ error $\sigma(P3/P0)$ that exceeds the \P3P0 or $w$ value by more than a factor of 3 are also considered as non-significant detections. For a more conservative factor of 1, hence taking values with $\sigma(P3/P0)>P3/P0$ or $\sigma(w)>w$ as non-significant detections, we find consistent results within the errors. For non-significant detections, we use upper limits (UL) in the analysis, where $UL=\sigma(P3/P0)+P3/P0_\mathrm{non-significant}$ for positive and $UL=\sigma(P3/P0)$ for negative corrected \P3P0 values. The definition is analogous for $w$. All presented \P3P0 and $w$ values are background and bias corrected.\\

During our discussion we will refer to different thresholds for \P3P0 and $w$ to divide the sample according to the dynamical state of the clusters. These dividing boundaries are taken from our previous work W13, where we also defined the significance $S$ of a \P3P0 or $w$ value as the ratio of the bias-corrected signal with respect to the obtained error. For high-quality data ($S>3$) we established two {\it morphological P3/P0 boundaries} to divide the sample into relaxed ($P3/P0 <10^{-8}$), mildly disturbed ($10^{-8}< P3/P0 <5\times10^{-7}$), and disturbed objects ($P3/P0 >5\times10^{-7}$). High $S$ values down to 10$^{-8}$ allow for this detailed classification. When dealing with low count observations, we reach $S=1$ around 10$^{-7}$ and use this value as {\it simple \P3P0 boundary} to separate disturbed and relaxed clusters. Owing to the data quality of the high-z samples (see Table \ref{statistics}), we only used the \P3P0 boundary at $10^{-7}$ for our analysis. \\
For the center shift parameter, we used $w=0.01$ to split the sample. Since $w$ is only severly affected by Poisson noise for considerably less than 1\,000 net photon counts within \r500 for a reasonably low background, this threshold can be used for high- and low-quality data.

\section{Data quality}
\label{DataQuality}

\begin{figure}
 \begin{center}
  \includegraphics[width=\columnwidth]{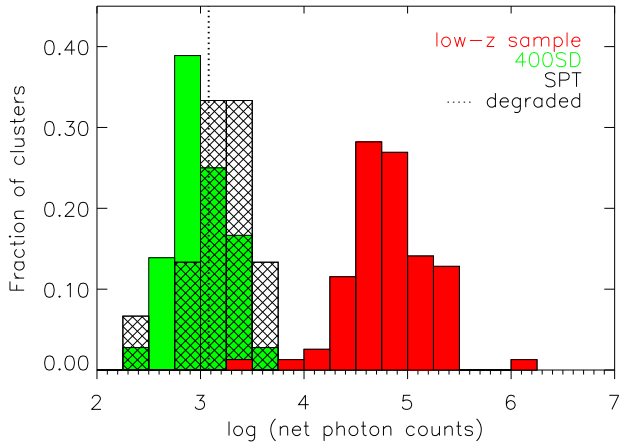}
 \end{center}
\caption{Overview of the net photon counts distribution within \r500 of the low-z (red filled histogram), high-z 400SD (green filled histogram) and high-z SPT sample (black dashed histogram). The dotted line indicates the net photon counts of the degraded data.}
\label{histo_cts}
\end{figure}

The strongest potential disadvantage when dealing with a combination of low- and high-z observations is the difference in the photon statistics of the observations, as can be seen by comparing Figs. \ref{lowzim} (left) and \ref{highzim}. Details of the sample statistics are given in Table \ref{statistics}, which shows that the low-z sample is not only larger in numbers but also in terms of higher photon statistics and a higher ratio of net (signal) to background photon counts (S/B). This results in a significant difference between the two samples in the extent and importance of photon shot noise. As we have shown in our previous work W13, photon shot noise can have a severe effect on the determination of the cluster morphology. We studied and quantified these effects and the influence of the background as a function of photon counts and S/B ratio for \P3P0 and $w$. We found that the center shift parameter can be determined with a small error even below the $w=0.01$ threshold for low photon statistics ($<1\,000$ net photon counts) and a reasonable S/B of e.g. $\sim2$. We can therefore obtain reasonable results for all morphologies, partly with relative large errors for very relaxed objects. The power ratio method needs sufficient photon counts to overcome the influence of Poisson noise, however. We showed that this problem is not important for disturbed objects, which do not suffer severly from shot noise and thus enable an accurate estimation even for low-quality data. For decreasing photon counts, however, mildly disturbed and relaxed objects undergo a boost of their signal due to an underestimation of the bias contribution that yields substructure parameters that are too high. In the case of excessive noise, we obtain a non-significant result. High-quality data therefore enable a more reliable determination of \P3P0 ($w$) and better statistics, including a higher number of clusters with $P3/P0>0$ ($w>0$) and a higher mean significance $\langle S \rangle$. A direct comparison between low- and high-quality data may thus not be conclusive. 

Fig. \ref{histo_cts} shows that the low-z data have more than sufficient photon counts with a mean of $\sim 97\,000$ net photon counts within \r500 to give \P3P0 and $w$ values with very good error properties and large $S$. The high-z objects, however, peak just above 1\,000 net photon counts with a mean of $\sim1\,200$ for 400SD and $\sim1700$ for SPT. According to simulations presented in W13, these high-z observations meet the criteria to roughly separate the sample into disturbed clusters with high and accurately determined substructure parameters and relaxed ones with parameters below the \P3P0 ($w$) threshold with large errors or non-significant detections. High-z observations contain a higher contribution from the background with a mean S/B of $\sim3.5$. This causes additional uncertainties due to the extra noise from the background and results in the low number of objects with $S>1$. To obtain conclusive results we need to establish the influence of noise and the possible boost of the \P3P0 ($w$) signal due to the lower data quality in the high-z sample. \newline

\subsection{Degrading of high-quality low-z observations}
\label{degrading}

To test how robust our results are to the difference in the data quality of the samples, we first performed our analysis using the high-quality or so-called undegraded low-z data. In addition, we created a degraded low-z sample by aligning the data quality of the low-z observations to that of the high-z objects. This was done by degrading the high-quality low-z observations to the photon statistics (1\,200 net photon counts and S/B=3.7 within \r500) of the 400SD high-z sample (see Table \ref{statistics}). The degrading was done in several steps, taking care of the different net and background photon counts and the increased Poisson noise. Two examples of degraded cluster images are given in Fig. \ref{lowzim} (right panels), compared with the undegraded images (left panels). The undegraded cluster image ($IM_{0}$) is not background subtracted. 
In the following recipe we denote images with capital letters and photon counts with lowercase letters. The recipe to obtain a low-z cluster  and background image with the same photon statistics as the average high-z cluster is outlined in steps 1-4. However, observations with low photon statistics do not only lack the sufficient number of photon counts, but also suffer from a considerable amount of Poisson noise. This is included by adding additional Poisson noise to the degraded image using the {zhtools}\footnote{hea-www.harvard.edu/RD/zhtools} task \textit{poisson}. In steps 5-7 we summarize the statistical analysis using the Poissonized realizations of these images.\\

\begin{enumerate} 
  \item Extract total photon counts ($im_{0}$) and background photon counts ($bkg_{0}$) within \r500 from the undegraded cluster ($IM_{0}$) and background image ($BKG_{0}$). Obtain net photon counts of the cluster as $cl_{0}=im_{0}-bkg_{0}$ and the S/B as $cl_{0}/bkg_{0}$.
  \item Calculate the additional background photon counts needed to obtain an S/B=3.7: $bkg_{add}=(cl_{0}/3.7)-bkg_{0}$. Rescale the undegraded background image by $bkg_{add}$: $BKG_{add}=BKG_{0}*bkg_{add}/bkg_{0}$
  \item Add the additional background image to the undegraded cluster image: $IM_{1}=IM_{0}+BKG_{add}$. This image has the desired S/B of 3.7.
  \item Rescale $IM_{1}$ to 1\,530 total photon counts within \r500: $IM_{deg}=IM_{1} * (1530/im_{1})$. Due to its S/B of 3.7, this degraded cluster image $IM_{deg}$ comprises 330 background and 1\,200 net photon counts. Rescale $BKG_{0}$ to 330 photon counts within \r500: $BKG_{deg}=BKG_{0} * (330/bkg_{0})$ to obtain the degraded background image. 
  \item Create 100 Poissonized realizations of the degraded cluster image $IM_{deg}$. Calculate background- and bias-corrected power ratios and center shifts including their errors for all 100 realizations of the cluster as described in W13.
  \item Randomly select one realization per cluster to create a new sample of 78 degraded low-z observations and obtain statistical measures like BCES fits or mean values.
  \item Repeat the previous step 100 times for statistical purposes and obtain the mean values. These are quoted when discussing our results including the mean errors.
\end{enumerate}


\section{Results}
\label{Results}

We studied the evolution of the substructure frequency up to $z=1.08$ using different statistical measures on the morphology estimators \P3P0 and $w$: i) fitting the data in the \P3P0-z and $w$-z plane with the linear relation $log(Y)=A\times log(z/0.25) + B$ for Y=\P3P0 and $w$ respectively, ii) calculating mean values for the different redshift intervals and iii) analyzing the fraction of relaxed and disturbed objects using \P3P0 and $w$ boundaries. For non-significant detections, we used upper limits as discussed in Sect. \ref{Sect3}. These are not included in the BCES fits given in Table \ref{Summary_asurv} and Figs. \ref{P3_orig}-\ref{original_data_w}. All analyses were performed on the log-distribution of \P3P0 and $w$ to take into account very low \P3P0 and $w$ values. Fitting parameters were calculated using the BCES (Y|X) fitting method \citep{Akritas1996}, which minimizes the residuals in Y. \\
To study the \P3P0-z and $w$-z relation we formed two samples to study possible selection effects of the high-z samples: i) sample I - low-z sample and high-z subsample of 400SD sample, ii) sample II - low-z sample and high-z subsample of SPT sample. We argue that using the degraded low-z data might be essential to obtain reliable and conclusive results. We therefore performed the identical analysis on sample I/II and the degraded sample I/II, where we used the degraded low-z data. We point out that only the high-quality low-z observations are degraded and thus are different in sample I/II and degraded sample I/II. The high-z data remains unchanged. In the following we focus on the heavily noise-affected \P3P0 parameter and then consider the more robust $w$ parameter. \\

During our analysis, we tried to include the information given by the upper limits in the \P3P0-z and $w$-z fits and tested the ASURV \citep{Feigelson1985,Isobe1986} and the LINMIX\_ERR \citep{Kelly2007} routine. For upper limits, both methods use estimated data points for fitting that are computed from the input upper limit and the distribution of the detected data points. Several tests using simulated images showed that the estimated data points are strongly coupled to the fit obtained from the detected data points and do not reflect the true \P3P0 values. Since the censorship in our data is due to low counts and dependent on \P3P0 itself, we conclude that our data do not fulfill the requirements for these routines to work properly.\\

\begin{table}[h!]
\begin{center}
\caption{Overview of the BCES (Y|z) fits in the log-log plane using the linear relation $log(Y)=A\times log(z/0.25) + B$ for Y=\P3P0 and $w$, respectively. Upper limits are omitted for these fits.}
\begin{tabular}{|l|cc|c|}
\hline & &  & \\[-1.5ex] 
\P3P0&  $A$ & $B$ & Fig. \\
\hline& & &    \\[-1.5ex] 
Sample I & 1.01$\pm$0.31 & -6.74$\pm$0.10  & \ref{P3_orig}, left\\
Sample II &  0.59$\pm$0.36 & -6.90$\pm$0.12 &  \ref{P3_orig}, left\\
Degraded sample I & 0.24$\pm$0.28 & -6.03$\pm$0.08& \ref{P3_deg}, left\\
Degraded sample II &  0.17$\pm$0.24 & -6.07$\pm$0.08 &\ref{P3_deg}, left \\
\hline\hline& & &  \\[-1.5ex]
$w$ &  $A$ & $B$ &  Fig. \\
\hline& & &   \\[-1.5ex]
Sample I & 0.18$\pm$0.14 & -2.01$\pm$0.04  &\ref{original_data_w}, left\\
Sample II & 0.02$\pm$0.13 & -2.05$\pm$0.04 & \ref{original_data_w}, left \\ 
Degraded sample I & 0.23$\pm$0.12 & -2.00$\pm$0.04 & \\
Degraded sample II & 0.07$\pm$0.11 & -2.04$\pm$0.04 & \\
\hline
\end{tabular}
\label{Summary_asurv}
\end{center}
\end{table}
\subsection{\P3P0-z relation}

\begin{figure*}
 \begin{center}
  \includegraphics[width=\columnwidth]{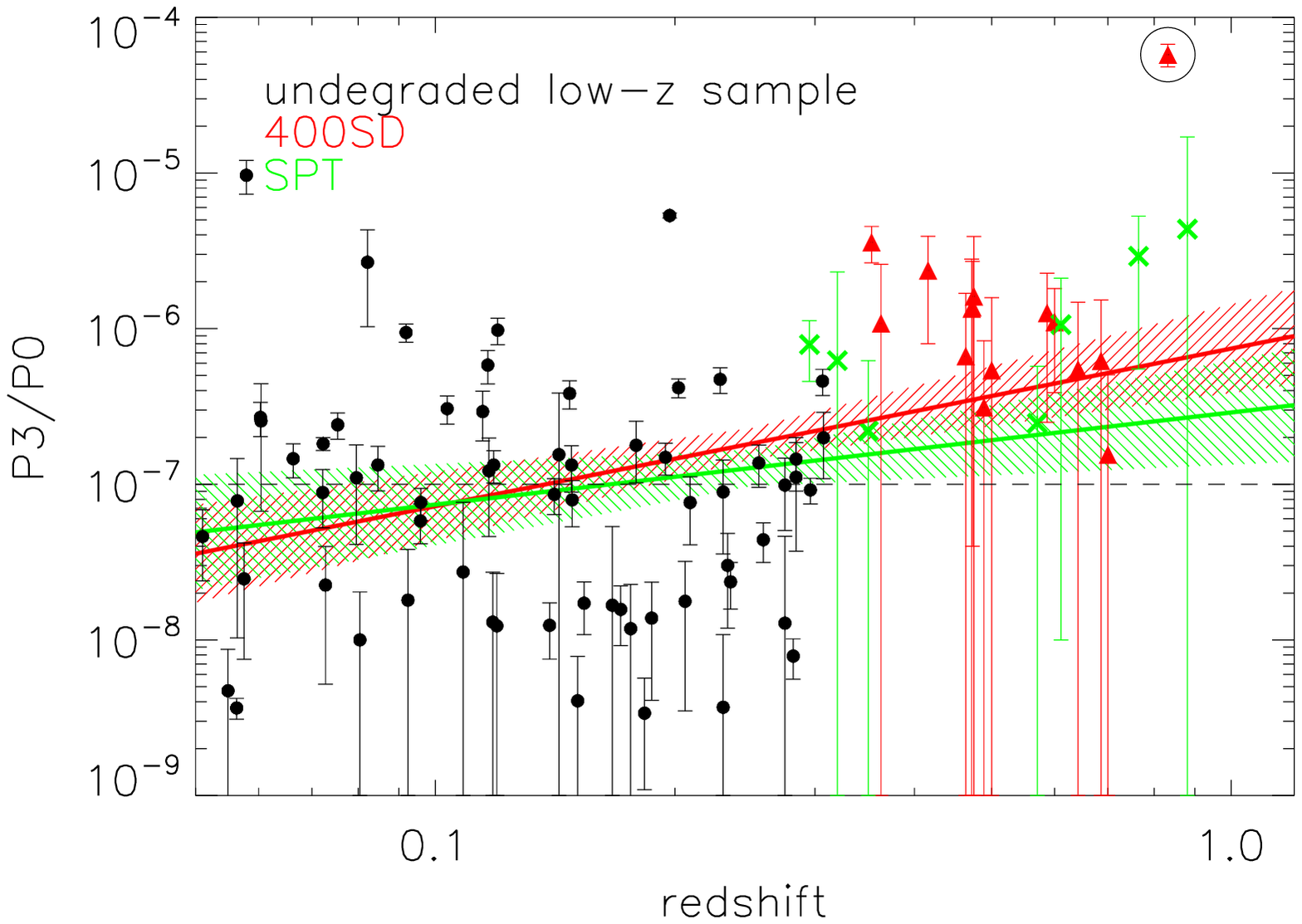}
  \includegraphics[width=\columnwidth]{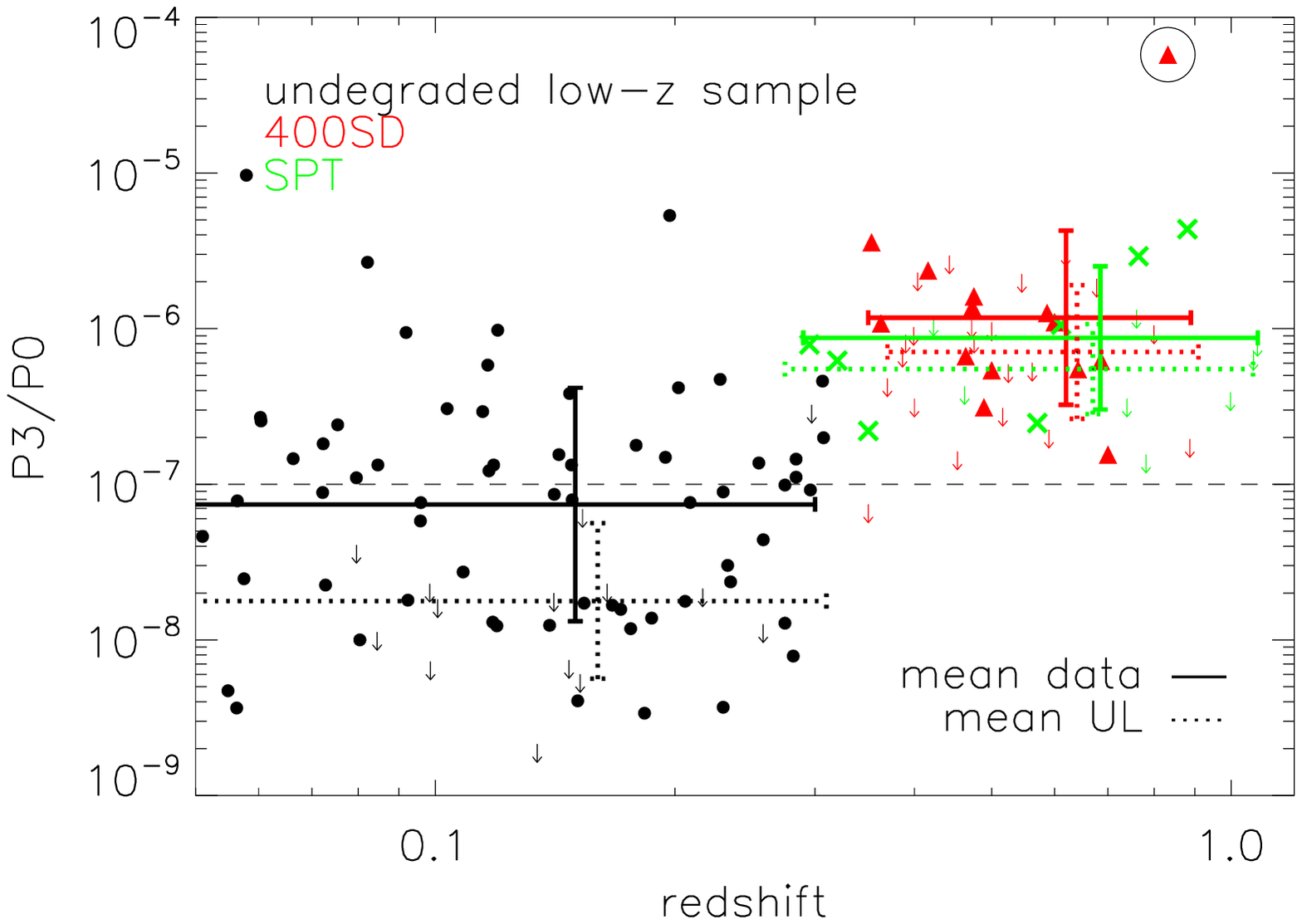}
 \end{center}
\caption{Undegraded \P3P0-z relation. Low-z (black circles), 400SD (red triangles), and SPT (green crosses) sample. Left: The BCES fit to sample I is shown as a red line while the green line indicates the fit to sample II. The dashed areas show the 1-$\sigma$ error of best-fitting values. Fitting parameters are given in Table \ref{Summary_asurv}. The very structured 400SD cluster 0152-1358 at $z\sim0.8$ with $P3/P0>10^{-5}$ is marked by a black circle. Excluding this cluster from sample I gives consistent results. In addition we show the \P3P0 boundary at $10^{-7}$ (dashed line). Right: Same data points as on the left, but including upper limits as downward arrows. For all three samples the solid lines give the mean of the log distribution of the significant data points including the 1-$\sigma$ errors, while the dotted lines show the mean of the upper limits.}
\label{P3_orig}
\end{figure*}

We first discuss the structure parameter \P3P0 as a function of redshift for sample I and II using Fig. \ref{P3_orig}. On the left side we show only the significant data points, while we include non-significant results as upper limits (arrows) on the right. For illustration, we show the \P3P0 boundary at $10^{-7}$ to separate relaxed and disturbed objects. When looking at this figure, one immediately notices the lack of significant detections of high-z clusters with $P3/P0<10^{-7}$. In addition, essentially all upper limits are found above this \P3P0 boundary. We quantified the \P3P0-z relation using the undegraded low-z data and different statistical measures. On the left of Fig. \ref{P3_orig} we show the linear BCES fit. For sample I we obtained a more than 3$\sigma$ significant slope with $A=1.01\pm0.31$, for sample II we found a somewhat shallower slope of $A=0.59\pm0.36$. We then tested the influence of the very structured 400SD cluster 0152-1358 ($z\sim0.8$) with $P3/P0>10^{-5}$ on the fit, finding a shallower, but consistent slope when excluding it from the fit. \\

Another way of quantifying the observed relation is computing the fraction of relaxed and disturbed objects in comparison to upper limits, which are shown in Table \ref{fractions}. Because of the high data quality of the undegraded low-z observations, the fraction of upper limits is small. All these objects can be considered as relaxed clusters because \P3P0 can detect significant signals well below $10^{-7}$ for such good data quality. Their non-significant signals or upper limits are consistent with $P3/P0 << 10^{-7}$. In addition, we find 45\% of the low-z objects to be relaxed. The majority of clusters in this sample is found below the \P3P0 threshold of $10^{-7}$ with a mean of the log \P3P0 distribution of $-7.1\pm0.8$. The high-z samples yield a higher mean of $-5.9\pm0.6$ ($-6.1\pm0.5$) for 400SD (SPT). The mean values are given in Table \ref{mean_median} and are denoted as $mean\ data$ for the significant data points and $mean\ UL$ for the upper limits. We plot the mean data values in Fig. \ref{P3_orig} on the right side to illustrate this offset. In addition, we add the mean UL values to emphasize again the difference in the location of upper limits for the high- and low-quality data.\\

\begin{figure*}
 \begin{center}
  \includegraphics[width=\columnwidth]{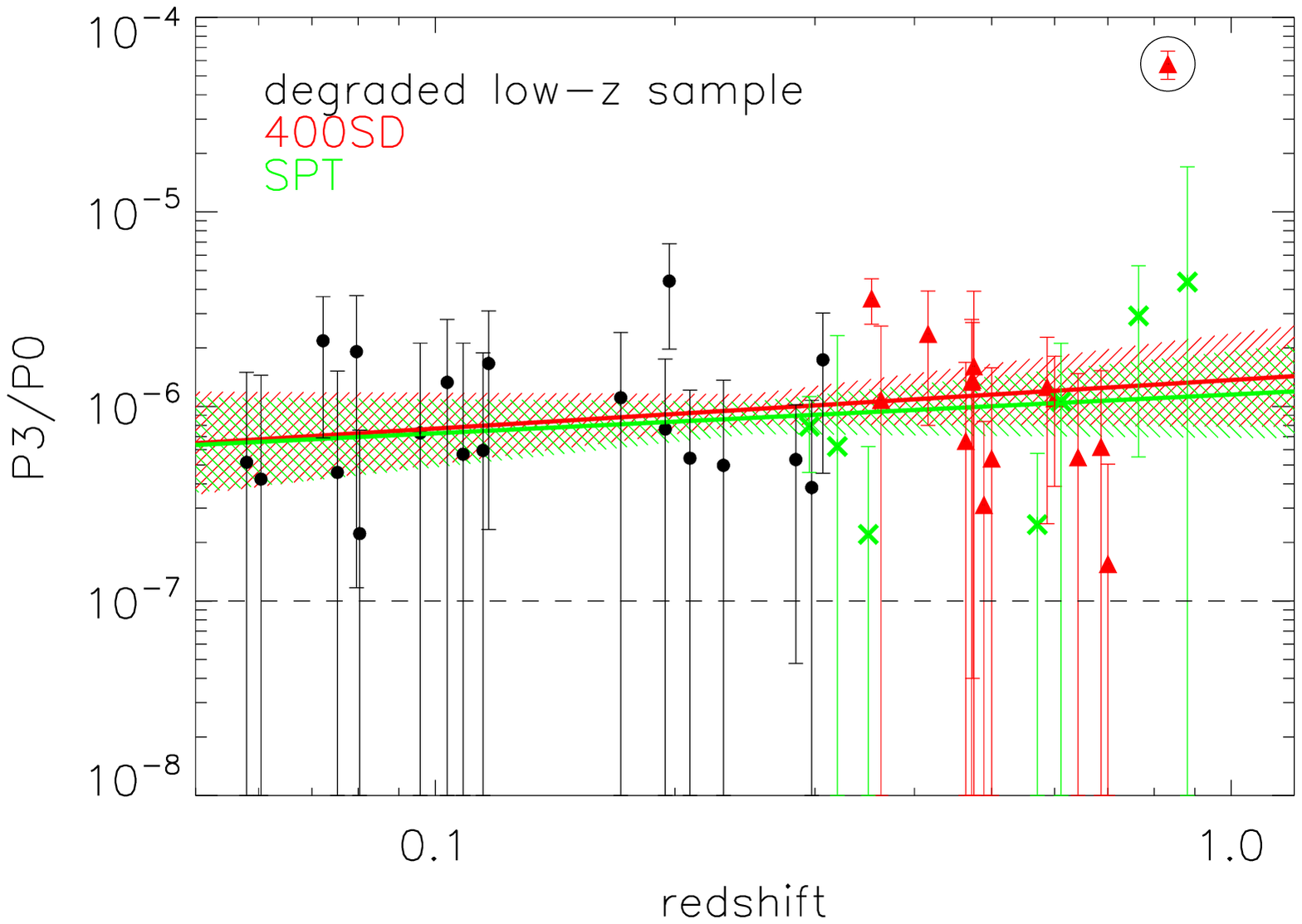}
  \includegraphics[width=\columnwidth]{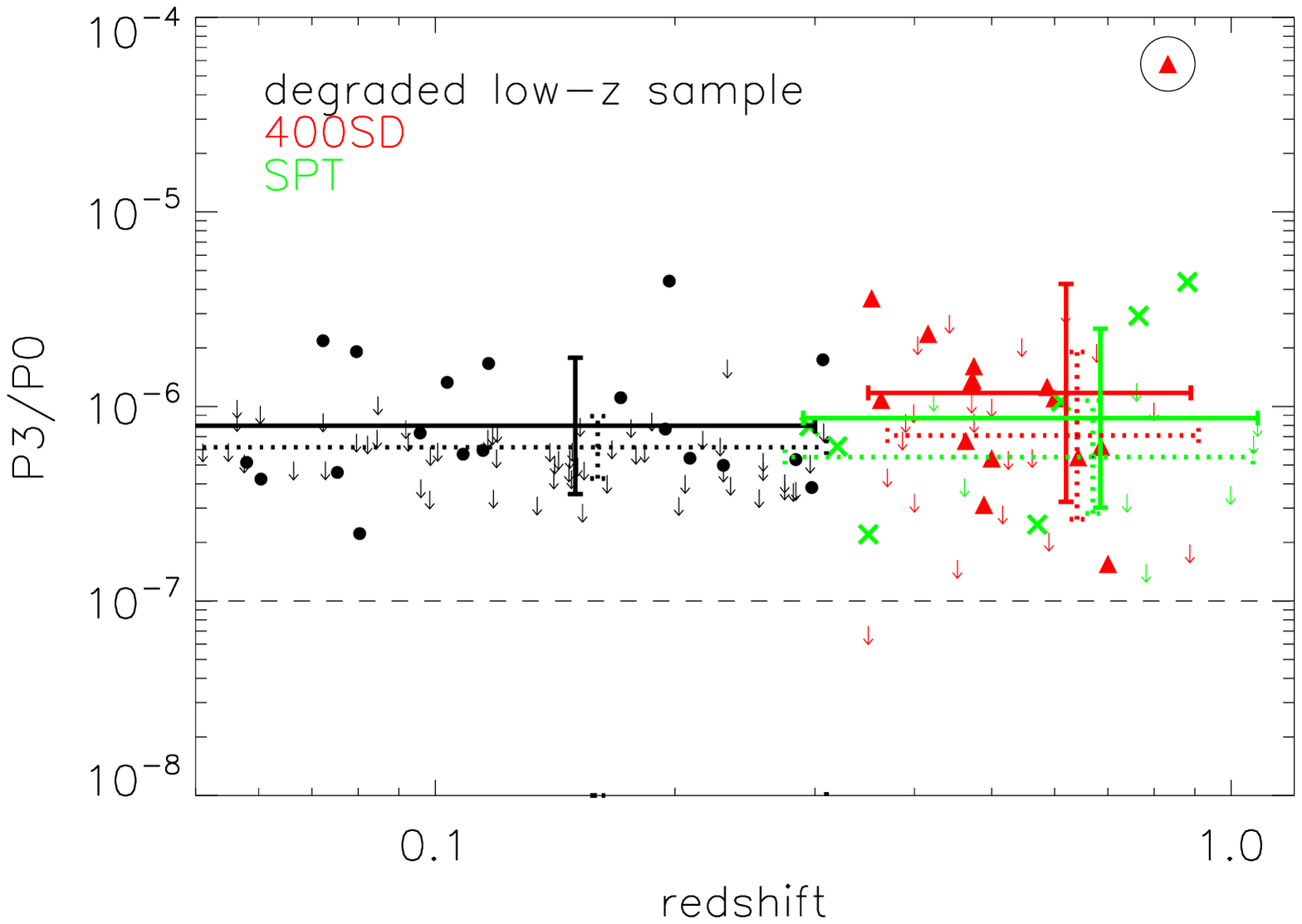}
 \end{center}
\caption{Degraded \P3P0-z relation. Details are the same as in Fig. \ref{P3_orig}.}
\label{P3_deg}
\end{figure*}

All statistical measures used on this dataset so far give a clear trend of a larger fraction of disturbed clusters at higher redshift. This conclusion should not be drawn without caution, however, since we are comparing very different datasets. We already argued that \P3P0 is heavily influenced by noise for observations with low net photon counts and/or high background. The computation of substructure parameters for the high-z objects therefore suffers severely from noise. According to results presented in W13, we can obtain significant \P3P0 values for the majority of the disturbed clusters even with fewer than $1\,000$ net photon counts within \r500. Mildly disturbed and relaxed objects will mostly either yield non-significant detections or undergo a boost of the \P3P0 signal. Except for some mildly disturbed objects whose \P3P0 values are just below the $10^{-7}$ boundary in the undegraded case, this boost will not result in $P3/P0>10^{-7}$. We should thus be able to very roughly separate the sample into disturbed ($P3/P0>10^{-7}$) and relaxed ($P3/P0<10^{-7}$ and upper limits) objects. \\
We repeated the analysis using the degraded low-z data and show the results in Fig. \ref{P3_deg}. We found significantly shallower slopes of $A=0.24\pm0.28$ ($A=0.17\pm0.24$) and higher intercepts $B$ for the degarded sample I (II). This is due to the apparent loss of data points with $P3/P0<10^{-7}$ and large errors on the detected \P3P0 signals after degrading. We find a significant increase of the upper limit fraction to 72\% while the fraction of relaxed clusters decreases from 45\% to on average 0\% (Table \ref{fractions}). The fraction of disturbed objects stays roughly the same, showing that we can detect a signal for the majority of structured objects while only a small number gives upper limits. With these low-quality data, we cannot measure a significant \P3P0 value for mildly disturbed or relaxed clusters anymore, but only detect disturbed objects. \\
For the high-z samples, we found no objects with $P3/P0<10^{-7}$ but a large number of upper limits (Table \ref{fractions}) and a disturbed cluster fraction of 42\% for 400SD and 47\% for SPT. Assuming that the majority of the disturbed objects yield significant detections, we found a slightly higher fraction of disturbed objects in the high-z samples than in the degraded low-z sample. \newline
We performed more tests by varying the degree of degradation of the low-z data. We found that the larger the disagreement between the net photon counts and S/B of the samples, the more biased the obtained slope or mean value. It is therefore of extreme importance to take this issue into account when analyzing the \P3P0-z relation.\\ 

 \begin{table}[h!]
 \centering
 \caption{Fraction of relaxed and disturbed objects using \P3P0 and $w$ boundaries taken from W13 (see Sect. \ref{Sect3}). Upper limits (UL) are given for non-significant detections.}
 \begin{tabular}{|c|c|c|c|}
 \hline& & &    \\[-1.5ex] 
 \P3P0 & \multicolumn{1}{|c|}{$P3/P0<10^{-7}$} & \multicolumn{1}{|c|}{$P3/P0>10^{-7}$} & \multicolumn{1}{|c|}{UL}\\
 \hline& & &  \\[-1.5ex]
Undeg. low-z data & 45\% & 37\% & 18\% \\
Deg. low-z data\tablefootmark{a} &  0\% & 31\% & 72\% \\
 400SD & 0\% & 42\% & 58\%\\
 SPT & 0\% & 47\% & 53\%  \\
\hline\hline& & &  \\[-1.5ex]
 $w$ & \multicolumn{1}{|c|}{$w<0.01$} & \multicolumn{1}{|c|}{$w>0.01$}& \multicolumn{1}{|c|}{UL}\\
 \hline& & & \\[-1.5ex]
Undeg. low-z data &  58\%  & 41\%  & 1\% \\
 Deg. low-z data\tablefootmark{a} &   52\% & 43\% & 6\% \\
 400SD & 39\% & 53\% & 8\%\\
 SPT & 60\% & 33\% & 7\% \\
 \hline
 \end{tabular}
 \label{fractions}
 \tablefoot{(a) mean values of 100 randomly selected samples.}
 \end{table}

\subsection{$w$-z relation}
\label{wz}

\begin{figure*}
 \begin{center}
  \includegraphics[width=\columnwidth]{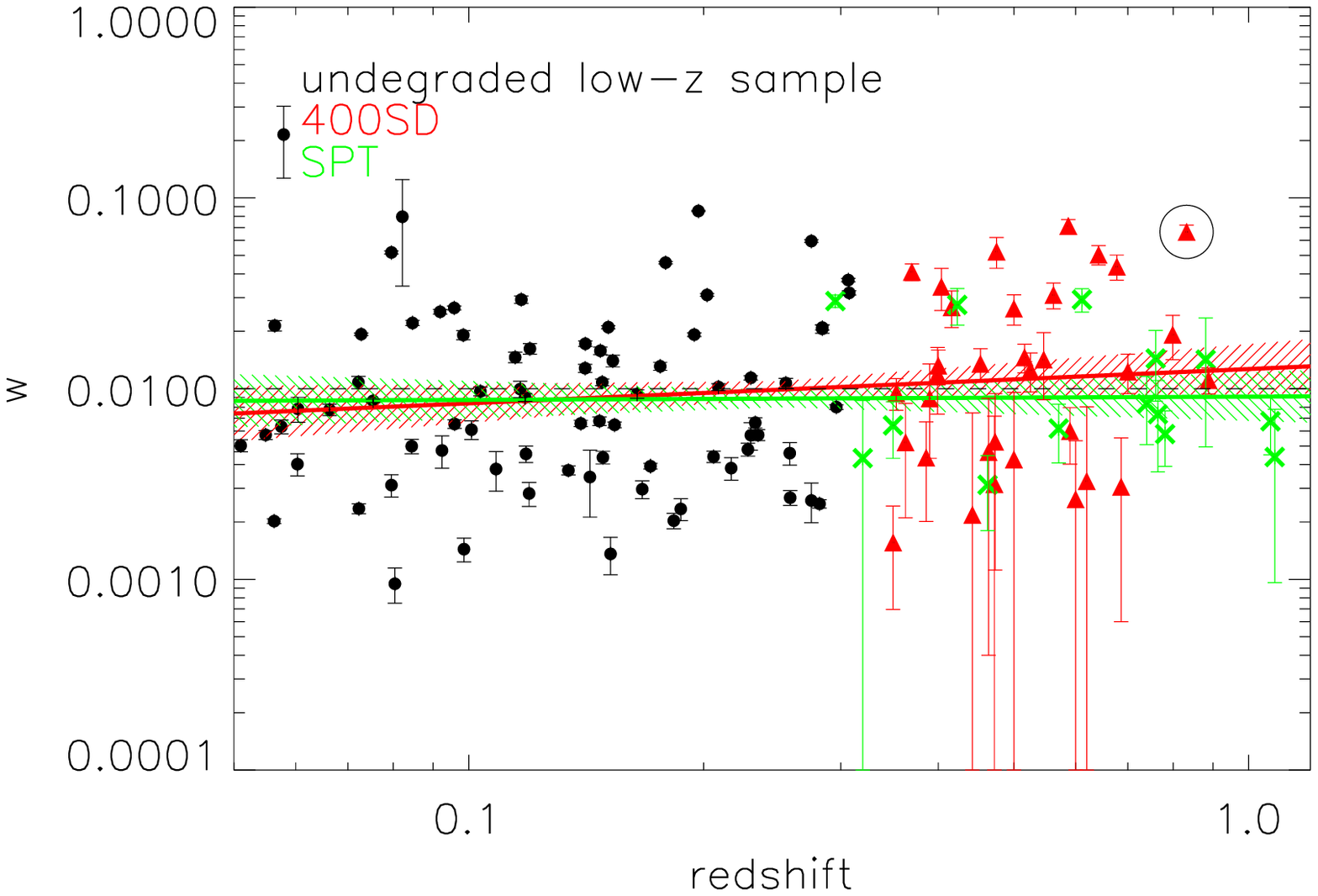}
  \includegraphics[width=\columnwidth]{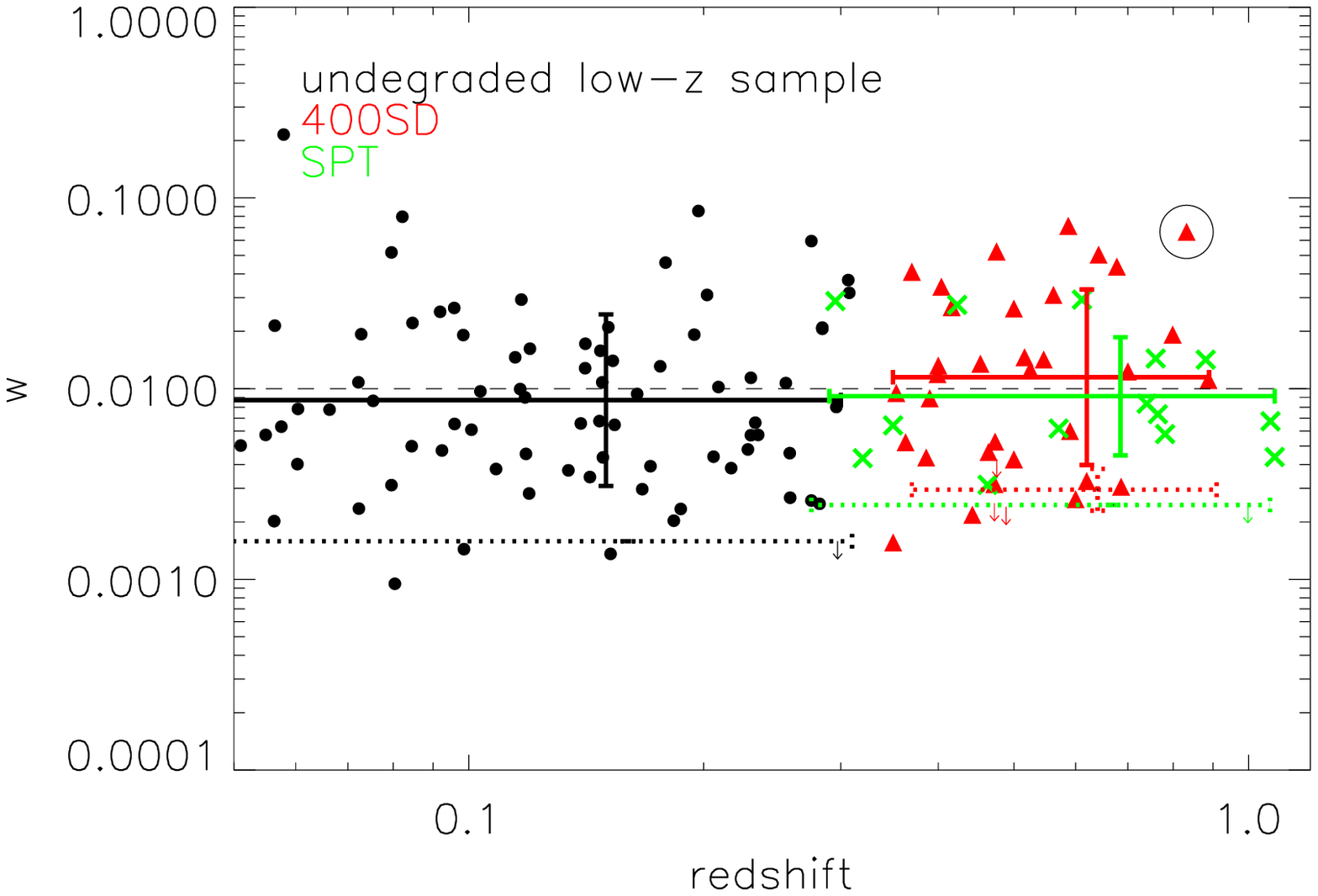}
 \end{center}
\caption{Undegraded $w$-z relation. Details are the same as in Fig. \ref{P3_orig}, except for the $w$ boundary at $10^{-2}$ (dashed line).}
\label{original_data_w}
\end{figure*}

Analogously to \P3P0, we used the same statistical measures on the $w$ parameter to probe its behavior as a function of redshift. Fig. \ref{original_data_w} shows the $w$ distribution for sample I and II, including upper limits on the right and the $w=0.01$ boundary to seperate relaxed and disturbed objects. We performed a linear BCES fit and give the fitting parameters in Table \ref{Summary_asurv}. The fits are illustrated on the left side of Fig. \ref{original_data_w}, with slope $A=0.18\pm0.14$ ($A=0.02\pm0.13$) for sample I (II). These slopes are both positive, but not significant and consistent with zero within 1-$\sigma$. In contrast to \P3P0, low- and high-z clusters populate the full $w$ range. This is reflected in the very similar mean values of the samples and their upper limits. We show these values in Table \ref{mean_median} and Fig. \ref{original_data_w} on the right side. \\
Because the $w$ parameter is not very sensitive to noise when dealing with $>1\,000$ net photon counts and a background that is not too high - as is the case with the high-z observations -, degrading the low-z observations to match the data quality of the 400SD clusters shows little effect. 
All statistical measures show very similar results when using the degraded low-z sample (Tables \ref{Summary_asurv}-\ref{mean_median}). The slopes stay well within the errors, and the mean data value does not change either. Only the mean upper limit value increases slightly, because the undegraded low-z data contains only one upper limit, but the degraded sample contains a few more. This is reflected in the slight increase of the upper limit fraction from 1\% to 6\%, which is very similar to those of the 400SD (8\%) and SPT (7\%) sample. The fraction of relaxed objects decreases slightly for the degraded data from 58\% to 52\%, while it increases for disturbed objects from 41\% to 43\%. These changes are within the errors and again show the robustness of $w$ against noise. Comapring the low-z fractions with those of the high-z samples, we see a very similar behavior of the SPT clusters, but the 400SD sample shows a larger fraction of objects with $w>0.01$. 

 \begin{table}[h!]
 \begin{center}
 \caption{Mean log(\P3P0) and log($w$) values for the low- and high-redshift samples. We give the mean of the significant data points (mean data) and the upper limits (mean UL) including their 1-$\sigma$ errors.}
 \begin{tabular}{|c|cc|c|c|}
\hline& & & &   \\[-1.5ex] 
  & \multicolumn{2}{|c|}{Low-z data} & 400SD & SPT \\ 
 \hline& & &  \multicolumn{2}{c|}{} \\[-1.5ex] 
 log(\P3P0) & undegraded & degraded & \multicolumn{2}{c|}{undegraded} \\ 
 \hline & & & & \\[-1.5ex]
 Mean data & -7.1$\pm$0.8& -6.1$\pm$0.4 & -5.9$\pm$0.6 & -6.1$\pm$0.5 \\
 Mean UL & -7.8$\pm$0.5 & -6.2$\pm$0.2 & -6.2$\pm$0.4 & -6.3$\pm$0.3 \\
 \hline\hline& & &  \multicolumn{2}{c|}{} \\[-1.5ex]
 log($w$) & undegraded & degraded & \multicolumn{2}{c|}{undegraded} \\
 \hline& & & &  \\[-1.5ex]
 Mean data& -2.1$\pm$0.5 &-2.1$\pm$0.4 &-1.9$\pm$0.5 &-2.0$\pm$0.3 \\
 Mean UL &-2.8 &-2.6$\pm$0.1 & -2.5$\pm$0.1& -2.6 \\
 \hline
 \end{tabular}
 \label{mean_median}
 \end{center}
 \end{table}

\section{Discussion}
\label{Discussion}

Assessing the dynamical state of a galaxy cluster calls for a well-studied method for detecting and quantifying substructure in the ICM. Well-understood error properties are of great importance, especially when dealing with high-z observations and thus low photon statistics. The two applied methods, power ratios and center shifts, fulfill these requirements. A strong correlation with a large scatter between \P3P0 and $w$ is known from previous studies \citep[e.g.][]{Boehringer2009} and therefore a similar trend in both relations is expected. Comparing the results obtained from applying \P3P0 and $w$ on sample I/II shows a very large discrepancy. While \P3P0 shows a significant increase with redshift in all statistical measures used, $w$ shows a positive but non-significant slope and no trend in the mean values either. We claim that the discrepancy between these results is caused by the inconsistent data quality of the full sample, which affects \P3P0 more than $w$. Taking into account the slopes, mean values, and fractions, one can conclude that $w$ is not sensitive to different data quality since the results hardly change. \newline
For of \P3P0, degrading the high-quality low-z observations to the net photon counts and background of the high-z objects yields very different results. The slope flattens significantly, yielding a similar result to $w$ - a positive but non-significant slope. The fraction of upper limits increases dramatically, because all relaxed objects yield non-significant detections. The fraction of low-z disturbed object is therefore only slightly smaller than those of the 400SD and SPT sample. Moreover, the mean data and mean UL values match those of the high-z samples when using equal data quality. \\

The results using \P3P0 and $w$ on this particular dataset show a similar trend. We found a very mild positive evolution, which is also consistent with no change with redshift within the significance limits. We excluded a strong increase of the disturbed cluster fraction with redshift and set an upper limit with the shallow slopes of the BCES fits. For the lower limit, we found no indication of a negative evolution because all statistical measures show an increase of \P3P0 and $w$ with redshift, but with low significance. \newline

\subsection{Comparison with previous studies}
\label{previousworks}

\begin{figure}
 \begin{center}
  \includegraphics[width=\columnwidth]{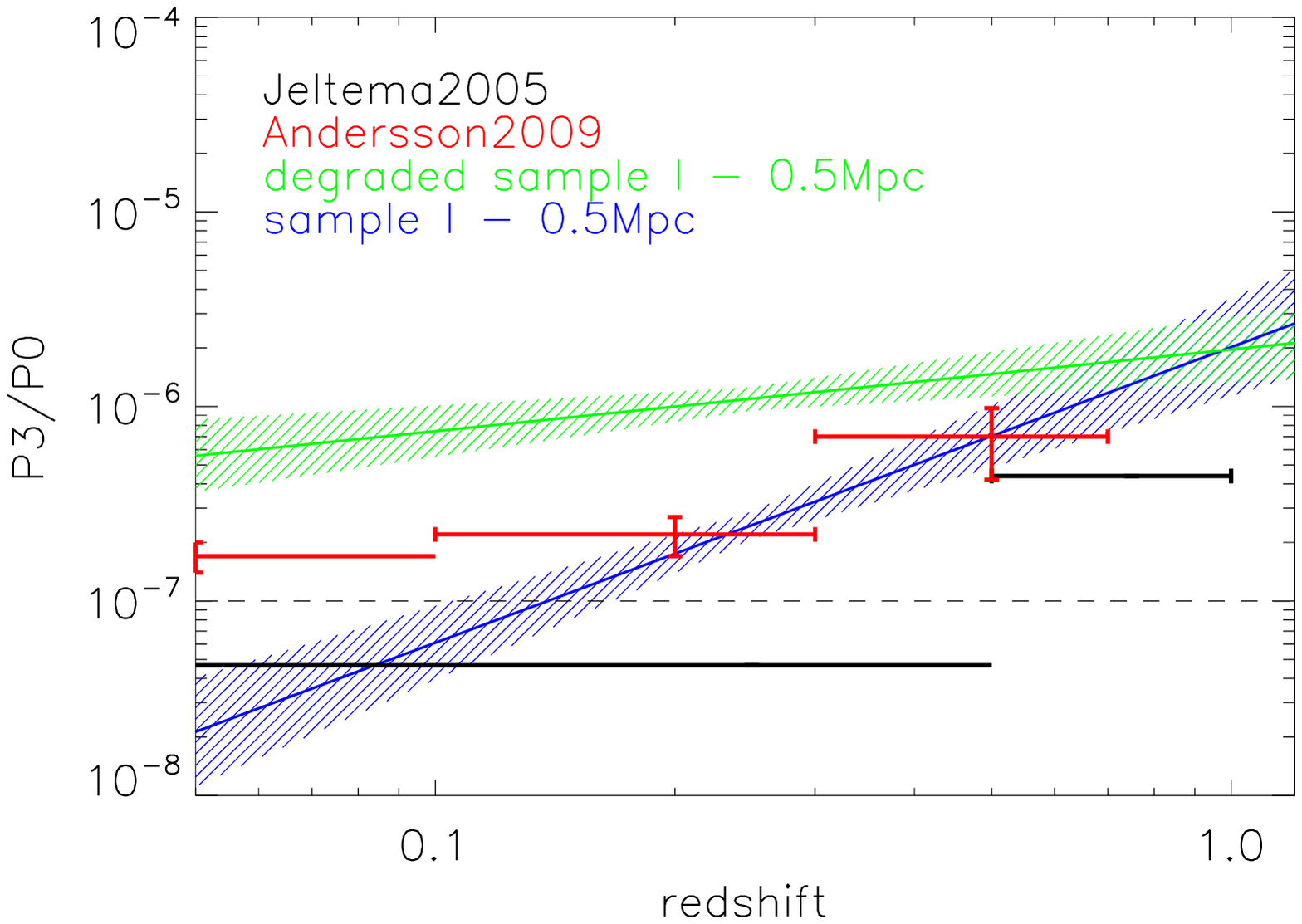}
 \end{center}
 \caption{Comparison with previous studies. From \citet{Jeltema2005} we show the mean \P3P0 values for $z<0.5$ and $z>0.5$ objects (black). Errors on these values are not provided. In addition, we plot the mean \P3P0 of \citet{Andersson2009} for three redshift bins ($0.069<z<0.1$, $0.1<z<0.3$ and $z>0.3$) in red. We provide our slope of the \P3P0-z plane calculated using an aperture of 0.5 Mpc for sample I (blue line) and the degraded sample I (green line) including the 1-$\sigma$ errors as dashed area. The dashed line indicates the \P3P0 boundary at $10^{-7}$.}
\label{previous}
\end{figure}

In the light of our finding that the different data quality between the low-z and high-z sample can severly bias the results, we compared our work with previous studies that did not take this problem into account. \citet{Jeltema2005} presented the first analysis of the \P3P0-z relation using 40 X-ray-selected luminous clusters in the $0.1<z<0.89$ range and a fixed physical aperture of 0.5 Mpc. They found the slope of the linear \P3P0-z relation to be $4.1\times10^{-7}$, but did not provide an intercept. We argue that a linear fit is not sensitive enough when working with a \P3P0 range of $10^{-9}$-$10^{-5}$. High \P3P0 values like that of the 400SD cluster 0152-1358 ($P3/P0>10^{-5}$ at z$\sim$0.8) dominate a linear fit, while low \P3P0 values are not adequately taken into account. We therefore did not include this result in Fig. \ref{previous}, which compares our findings with previous studies. 
\citet{Jeltema2005} also provided mean \P3P0 values for $z<0.5$ and $z>0.5$ objects. For a fair comparison, we calculated \P3P0 in the same 0.5 Mpc aperture, since \r500 is typically larger than 1 Mpc for the low-z sample and on average 0.8 Mpc for high-z objects. A fixed aperture of 0.5 Mpc probes the cluster structure on a different scale than \r500. For the 0.5 Mpc aperture, the slopes of the fits are steeper and the intercepts higher with $A=1.52\pm0.30$ ($A=1.08\pm0.39$) and $B=-6.61\pm0.07$ ($B=-6.75\pm0.10$) for sample I (II). After degrading, the \P3P0-z fits flatten significantly to $A=0.42\pm0.18$ ($A=0.08\pm0.31$) with $B=-5.96\pm0.06$ ($B=-6.09\pm0.10$) for the degraded sample I (II) and agree well with the degraded results when using \r500 as aperture. The general impression of a very mild increase of the disturbed cluster fraction with redshift thus holds also for the 0.5 Mpc aperture.
We show the fits for sample I and the degraded sample I using the 0.5 Mpc aperture in Fig. \ref{previous}. The discrepancy between our fit of the degraded sample I and the mean values of \citet{Jeltema2005} is apparent. While \citet{Jeltema2005} took general noise properties into account, they did not address the problem of the data quality difference, which results in an overestimation of the slope and a large offset between the mean low-z and high-z sample. \\
Another study was performed by \citet{Andersson2009}, who also calculated \P3P0 in an 0.5 Mpc aperture for 101 galaxy clusters in the range $0.07<z<0.89$. They reported an increase in \P3P0 and provided average \P3P0 values given for three redshift bins ($0.069<z<0.1$, $0.1<z<0.3$ and $z>0.3$). We see an offset to our degraded fits here as well.\\

Several studies using both simulations \citep[e.g.][]{Ho2006} and observations \citep[e.g.][]{Maughan2008, Plionis2002,Melott2001} explored the evolution of ellipticity with redshift. The asymmetry in the X-ray surface brightness distribution was studied by \citet{Hashimoto2007a}, reporting no significant difference regarding ellipticity and off-center between the low- and high-z sample, but a hint of a weak evolution for the concentration and asymmetry parameter. Recently, \citet{Mann2012} presented a study of the evolution of the cluster merger fraction using 108 of the most X-ray-luminous galaxy clusters at $0.15<z<0.7$. They used optical and X-ray data and classified mergers according to their morphological class, X-ray centroid - BCG separation and X-ray peak - BCG separation. They reported an increase of the fraction of disturbed clusters with redshift, starting around $z=0.4$.\\
In addition to observational studies, we compared our findings with those of \citet{Jeltema2008}, who studied the evolution of cluster structure with \P3P0 and $w$ in hydrodynamical simulations performed with Enzo, a hybrid Eulerian adaptive mesh refinement/N-body code. Their simulations did not include the effect of noise or instrumental response, therefore only a broad comparison to low-z observed data with high singal-to-noise is possible. They reported a dependence of the evolution of \P3P0 with redshift on the selection criterium and on the radius chosen. While for $w$ they found a significant increase with redshift for a mass as well as a luminosity cut, \P3P0 showed an evolution only for a luminosity-limited sample. In agreement with our results, they stated that the evolution of cluster structure is mild compared with the variety of cluster morphologies seen at all redshifts.\\

\subsection{Effect of cool cores}
\label{effectcoolcores}

Several studies showed that cool cores are preferentially found in relaxed systems. \citet{Santos2010a} and \citet{Andersson2009} found a negative evolution of the fraction of cool-core clusters, reporting that the number of cooling core clusters appears to decrease with redshift. This suggests a higher fraction of relaxed clusters at low than at high redshift. They also argued that the evolution is significantly less pronounced than previously claimed. \citet{Bauer2005} used the high-z end of the BCS sample and concluded that the fraction of cool-cores does not significantly evolve up to $z\sim0.4$.\\

It is therefore an interesting exercise to study whether the \P3P0-z and $w$-z relation is driven by the presence of a cool core or by the overall dynamical state of the cluster. To do this, we excluded the 0.1 \r500 region when calculating the centroid, but kept it to determine the X-ray peak. For an aperture of \r500 we found very similar slopes for both relations. For $w$ the slope becomes somewhat shallower but remains well within the 1-$\sigma$ error with $A=0.10\pm0.14$ ($A=-0.07\pm0.13$) and $B=-1.97\pm0.04$ ($B=-2.02\pm0.04$). As expected, degrading has no real effect on the center-excised $w$-z relation, and the mean values also stay well within the errors. Contrary to our findings, \citet{Maughan2008} reported a significant absence of relaxed clusters at high redshift using a sample of 115 galaxy clusters in the $0.1<z<1.3$ range and center shifts with the central 30 kpc excised as morphology estimator. \\
\P3P0, on the other hand, yields slightly higher values on averge when excluding the center, which results in a very similar slope of $A=0.97\pm0.29$ ($A=0.58\pm0.32$), but in a higher intercept of $B=-6.53\pm0.09$ ($B=-6.68\pm0.10$) and higher mean values for sample I (II). The same effects are seen when using the degraded low-z sample. The analysis was repeated using the fixed 0.5 Mpc aperture. We found a larger difference between the core-included and excised \P3P0-z relation for this aperture, because it is more sensitive to substructure in the inner region of the cluster. The obtained results are comparable with the \r500 case, however. We conclude that based on the method to obtain \P3P0 and $w$, the \P3P0-z and $w$-z relations seem to be mainly driven by the dynamical state of the cluster on the scale of the aperture radius.\\ 


\section{Conclusions}
\label{Conclusion}

We studied the evolution of the substructure frequency by comparing a merged sample of 78 low-z observations of galaxy clusters with the high-z subsample of the 400SD and SPT sample. The analysis was performed on two samples individually to exclude possible selection effects of the high-z samples: i) sample I: 78 low-z and 36 400SD, ii) sample II: 78 low-z and 15 SPT clusters. Power ratios \P3P0 and the center shift parameter $w$ were used to quantify the amount of substructure in the cluster X-ray images. \\

We found that directly comparing high-quality low-z and low-quality high-z observations using \P3P0
\begin{itemize}
 \renewcommand{\labelitemi}{$\bullet$}
 \item yields a very steep \P3P0-z relation with slopes of $1.01\pm0.31$ ($0.59\pm0.36$) for sample I (II),
 \item gives a significant difference in the mean \P3P0 values of the low-z and high-z samples, and
 \item returns a very large fraction of relaxed objects at low-z (45\%), but none at high-z.
\end{itemize}

However, as was shown in our previous work \citep[][]{Weissmann2013}, \P3P0 is very sensitive to noise and thus to the depth and quality of the observation. We corrected for the noise bias, but uncertainties in the results of low-quality data remained. Since there is a significant difference in the data quality of the samples, this problem needed to be considered during the analysis. 
We therefore degraded the high-quality low-z observations to the photon statistics of the high-z 400SD observations. This enabled a comparison of data with similar quality and thus more reliable results. Using equal data quality and \P3P0, we found
\begin{itemize}
 \renewcommand{\labelitemi}{$\bullet$}
 \item a weak, but not very significant evolution in the \P3P0-z relation with slopes of $0.24\pm0.28$ ($0.17\pm0.24$) for the degraded sample I (II),
 \item no difference in the mean \P3P0 value of the low-z and high-z samples,
 \item that all relaxed ($P3/P0<10^{-7}$) low-z clusters yield non-significant detections after degradation, and
 \item a slightly larger fraction of disturbed clusters in the high-z samples (42\% for 400SD, 47\% for SPT) than in the degraded low-z sample (31\%).
\end{itemize}

We performed the same analysis using the center shift parameter $w$ as morphology estimator. $w$ is more robust against Poisson noise and not very sensitive to the data quality difference of the samples. We therefore found very similar results using the undegraded and degraded low-z data, namely
\begin{itemize}
\renewcommand{\labelitemi}{$\bullet$}
 \item a very shallow slope of the $w$-z relation: $0.18\pm0.14$ ($0.02\pm0.13$) for sample I (II), $0.23\pm0.12$ ($0.07\pm0.11$) for the degraded sample I (II),
 \item no difference in the mean $w$ value of the low-z and high-z samples, and
 \item no significant difference in the fraction of relaxed and disturbed objects.
\end{itemize}

Considering that the 400SD high-z sample may contain an unrepresentatively large number of disturbed clusters, the slopes obtained using this dataset should be taken as upper limits. They are consistent with the results when using the SPT clusters as high-z sample, however. In summary, we agree with previous findings, which indicate an evolution of the substructure frequency with redshift. \\

We conclude that the results using \P3P0 and $w$ on this particular dataset show a similar and very mild positive evolution of the substructure frequency with redshift. However, within the significance limits, our findings are also consistent with no evolution. A strong increase of the disturbed cluster fraction is excluded and the BCES fits are taken as upper limits. For the lower limit, we found no indication of a negative evolution. All statistical measures show a slight increase of \P3P0 and $w$ with redshift, but with low significance. Larger samples of deep observations of $z>0.3$ galaxy clusters would provide a better way to quantify these relations and allow unambiguous conclusions.


\begin{acknowledgements}
We would like to thank the anonymous referee for constructive comments and suggestions. This work is based on observations obtained with XMM-\emph{Newton} and CHANDRA. XMM-\emph{Newton} is an ESA science mission with instruments and contributions directly funded by ESA Member States and NASA. The XMM-\emph{Newton} project is supported by the Bundesministerium f\"{u}r Wirtschaft und Technologie/Deutsches Zentrum f\"{u}r Luft- und Raumfahrt (BMWI/DLR, FKZ 50 OX 0001), the Max-Planck Society and the Heidenhain-Stiftung. A part of the scientific results reported in this article is based on data obtained from the Chandra Data Archive. AW acknowledges the support from and participation in the International Max-Planck Research School on Astrophysics at the Ludwig-Maximilians University. GC acknowledges the support from Deutsches Zentrum f\"{u}r Luft- und Raumfahrt (DLR) with the program ID 50 R 1004. HB and GC acknowledge support from the DfG Transregio Program TR33 and the Munich Excellence Cluster ``Structure and Evolution of the Universe''. 
\end{acknowledgements}

\bibliographystyle{aa}
\bibliography{library_paper2}

\begin{center}
\begin{longtable}{lcccccccc}
\caption{Details of the individual galaxy clusters including structure parameters.}\\
\hline\\[-1.2ex]
\multicolumn{1}{c}{Cluster} & \multicolumn{1}{c}{Redshift} & \multicolumn{1}{c}{\r500} & \multicolumn{1}{c}{\P3P0} & \multicolumn{1}{c}{UL}& \multicolumn{1}{c}{$w$} & \multicolumn{1}{c}{UL} & \multicolumn{1}{c}{Reference} & \multicolumn{1}{c}{Image}\\ [0.5ex]
\hline\\ [-1.8ex]
\endfirsthead
\caption{continued.}\\
\hline\\[-1.2ex]
\multicolumn{1}{c}{Cluster} & \multicolumn{1}{c}{Redshift} & \multicolumn{1}{c}{\r500} & \multicolumn{1}{c}{\P3P0} & \multicolumn{1}{c}{UL}& \multicolumn{1}{c}{$w$} & \multicolumn{1}{c}{UL} & \multicolumn{1}{c}{Reference} & \multicolumn{1}{c}{Image}\\ [0.5ex]
\hline\\[-1.2ex]
\endhead
\hline\\[-1.8ex]
\endfoot 
\hline \multicolumn{9}{|c|}{} \\[-1.5ex] 
\multicolumn{9}{|c|}{low-z sample} \\[0.5ex]
\hline\\[-1.2ex]
\object{RXCJ0307.0-2840}&0.26&1.18&$1.26\times10^{-8}$	       &1&$(4.59\pm0.52)\times10^{-3}$&0&1,2	&X	\\
\object{RXCJ0528.9-3927}&0.28&1.12&$(1.11\pm0.74)\times10^{-7}$&0&$(2.06\pm0.11)\times10^{-2}$&0&1,2	&X	\\
\object{RXCJ0532.9-3701}&0.28&1.23&$(9.88\pm4.84)\times10^{-8}$&0&$(2.59\pm0.61)\times10^{-3}$&0&1,2	&X	\\
\object{RXCJ0658.5-5556}&0.30&1.46&$(9.19\pm1.72)\times10^{-8}$&0&$(8.01\pm0.23)\times10^{-3}$&0&1*,2,3,5&X	\\
\object{RXCJ0945.4-0839}&0.15&1.06&$6.92\times10^{-8}$	       &1&$(1.40\pm0.10)\times10^{-2}$&0&1	&X	\\
\object{RXCJ2129.6+0005}&0.24&1.12&$(2.36\pm0.79)\times10^{-8}$&0&$(5.72\pm0.37)\times10^{-3}$&0&1	&X	\\
\object{RXCJ2308.3-0211}&0.30&1.20&$3.24\times10^{-7}$	       &1&$1.58\times10^{-3}$		&1&1*,2	&X	\\
\object{RXCJ2337.6+0016}&0.28&1.21&$(1.28\pm3.35)\times10^{-8}$&0&$(5.94\pm0.10)\times10^{-2}$&0&1*,2	&X	\\
\object{A68} 		&0.26&1.20&$(1.37\pm0.42)\times10^{-7}$&0&$(1.07\pm0.06)\times10^{-2}$&0&1*,3	&X	\\
\object{A115}		&0.20&1.13&$(5.33\pm0.19)\times10^{-6}$&0&$(8.54\pm0.05)\times10^{-2}$&0&1	&X	\\
\object{A209}		&0.21&1.22&$(7.63\pm3.55)\times10^{-8}$&0&$(1.02\pm0.04)\times10^{-2}$&0&1*,3	&X	\\
\object{A267}		&0.23&1.11&$(8.93\pm5.37)\times10^{-8}$&0&$(5.70\pm0.94)\times10^{-3}$&0&1	&X	\\
\object{A383}		&0.19&0.97&$(1.38\pm0.97)\times10^{-8}$&0&$(2.34\pm0.31)\times10^{-3}$&0&1*,3	&X	\\
\object{A773}		&0.22&1.32&$2.14\times10^{-8}$	       &1&$(3.83\pm0.52)\times10^{-3}$&0&1*,3	&X	\\
\object{A963}		&0.21&1.16&$(1.77\pm1.42)\times10^{-8}$&0&$(4.40\pm0.30)\times10^{-3}$&0&1	&X	\\
\object{A1413}		&0.14&1.21&$(1.55\pm2.31)\times10^{-7}$&0&$(3.44\pm1.32)\times10^{-3}$&0&1*,3,4,5&X	\\
\object{A1763}		&0.23&1.07&$(4.72\pm0.89)\times10^{-7}$&0&$(4.80\pm0.37)\times10^{-3}$&0&1	&X	\\
\object{A1914}		&0.17&1.40&$(1.57\pm0.65)\times10^{-8}$&0&$(3.92\pm0.13)\times10^{-3}$&0&1*,3,5&X	\\
\object{A2390}		&0.23&1.59&$(3.01\pm1.82)\times10^{-8}$&0&$(6.63\pm0.40)\times10^{-3}$&0&1	&X	\\
\object{A2667}		&0.23&1.19&$(3.69\pm7.12)\times10^{-9}$&0&$(1.14\pm0.03)\times10^{-2}$&0&1*,3	&X	\\
\object{A2204}		&0.15&1.37&$6.03\times10^{-9}$	       &1&$(1.36\pm0.30)\times10^{-3}$&0&1*,3,4,5&X	\\
\object{A2218}		&0.18&1.19&$(1.18\pm1.09)\times10^{-8}$&0&$(1.31\pm0.06)\times10^{-2}$&0&1*,3,5&X	\\
\object{RXCJ0232.2-4420}&0.28&1.12&$(1.45\pm0.55)\times10^{-7}$&0&$(2.09\pm0.06)\times10^{-2}$&0&1*,2	&X	\\
\object{A13}		&0.10&0.95&$(3.06\pm0.63)\times10^{-7}$&0&$(9.68\pm0.51)\times10^{-3}$&0&3	&X	\\
\object{A520}		&0.19&1.31&$(1.49\pm0.35)\times10^{-7}$&0&$(1.92\pm0.04)\times10^{-2}$&0&3	&X	\\
\object{A665}		&0.18&1.32&$(1.78\pm0.76)\times10^{-7}$&0&$(4.58\pm0.07)\times10^{-2}$&0&3*,5	&X	\\
\object{A1068}		&0.15&0.99&$7.41\times10^{-9}$	       &1&$(6.76\pm0.35)\times10^{-3}$&0&3,4*,5&X	\\
\object{A1589}		&0.07&0.88&$(8.85\pm3.60)\times10^{-8}$&0&$(1.08\pm0.08)\times10^{-2}$&0&3	&X	\\
\object{A2163}		&0.20&1.85&$(4.17\pm0.58)\times10^{-7}$&0&$(3.10\pm0.05)\times10^{-2}$&0&3	&X	\\
\object{A2717}		&0.05&0.74&$(4.62\pm2.23)\times10^{-8}$&0&$(5.03\pm0.36)\times10^{-3}$&0&3,4*,5&X	\\
\object{A3112}		&0.07&0.98&$(1.82\pm0.17)\times10^{-7}$&0&$(2.35\pm0.14)\times10^{-3}$&0&3	&X	\\
\object{A3827}		&0.10&1.21&$(7.62\pm1.78)\times10^{-8}$&0&$(6.53\pm0.34)\times10^{-3}$&0&3	&X	\\
\object{A3911}		&0.10&1.15&$(5.81\pm1.67)\times10^{-8}$&0&$(2.65\pm0.06)\times10^{-2}$&0&3	&X	\\
\object{A3921}		&0.09&1.08&$(9.44\pm1.27)\times10^{-7}$&0&$(2.53\pm0.07)\times10^{-2}$&0&3	&X	\\
\object{E1455+2232}	&0.26&0.97&$(4.40\pm1.25)\times10^{-8}$&0&$(2.68\pm0.24)\times10^{-3}$&0&3	&X	\\
\object{PKS0745-19}	&0.10&1.37&$7.24\times10^{-9}$	       &1&$(1.44\pm0.51)\times10^{-3}$&0&3,4*	&X	\\
\object{Sersic159-3}	&0.06&0.68&$(3.65\pm0.56)\times10^{-9}$&0&$(2.02\pm0.01)\times10^{-3}$&0&3	&X	\\
\object{ZW3146}		&0.28&1.21&$(7.87\pm2.28)\times10^{-9}$&0&$(2.49\pm0.13)\times10^{-3}$&0&3	&X	\\
\object{A2597}		&0.08&0.89&$(1.00\pm1.03)\times10^{-8}$&0&$(9.48\pm1.98)\times10^{-4}$&0&3,4*	&X	\\
\object{A1775}		&0.08&0.91&$(2.41\pm0.46)\times10^{-7}$&0&$(8.63\pm0.18)\times10^{-3}$&0&3	&X	\\
\object{A1837}		&0.07&0.79&$(1.46\pm0.36)\times10^{-7}$&0&$(7.76\pm0.53)\times10^{-3}$&0&3*,5	&X	\\
\object{RXCJ0014.3-3022}&0.31&1.41&$(4.60\pm0.88)\times10^{-7}$&0&$(3.71\pm0.08)\times10^{-2}$&0&2	&X	\\
\object{RXCJ1131.9-1955}&0.31&1.33&$(1.99\pm0.90)\times10^{-7}$&0&$(3.18\pm0.08)\times10^{-2}$&0&2	&X	\\
\object{A1651}		&0.08&1.21&$1.22\times10^{-8}$	       &1&$(4.99\pm0.43)\times10^{-3}$&0&5	&X	\\
\object{A133}		&0.06&0.92&$(2.47\pm1.72)\times10^{-8}$&0&$(6.32\pm0.53)\times10^{-3}$&0&3	&X	\\
\object{A2626}		&0.05&0.75&$(4.71\pm3.98)\times10^{-9}$&0&$(5.72\pm0.32)\times10^{-3}$&0&3	&X	\\
\object{A2065}		&0.07&1.05&$(2.25\pm1.73)\times10^{-8}$&0&$(1.93\pm0.03)\times10^{-2}$&0&3	&X	\\
\object{RXCJ0003.8+0203}&0.09&0.91&$(1.80\pm2.01)\times10^{-8}$&0&$(4.74\pm0.91)\times10^{-3}$&0&6	&X	\\
\object{RXCJ0006.0-3443}&0.11&1.05&$(2.93\pm1.03)\times10^{-7}$&0&$(1.46\pm0.09)\times10^{-2}$&0&6	&X	\\
\object{RXCJ0020.7-2542}&0.14&1.11&$2.00\times10^{-8}$	       &1&$(1.28\pm0.06)\times10^{-2}$&0&6	&X	\\
\object{RXCJ0049.4-2931}&0.11&0.80&$(2.73\pm4.93)\times10^{-8}$&0&$(3.79\pm0.89)\times10^{-3}$&0&6	&X	\\
\object{RXCJ0145.0-5300}&0.12&1.11&$(1.22\pm0.76)\times10^{-7}$&0&$(2.93\pm0.13)\times10^{-2}$&0&6	&X	\\
\object{RXCJ0211.4-4017}&0.10&0.64&$1.82\times10^{-8}$	       &1&$(6.09\pm0.69)\times10^{-3}$&0&6	&X	\\
\object{RXCJ0225.1-2928}&0.06&0.72&$(2.55\pm1.88)\times10^{-7}$&0&$(7.83\pm1.17)\times10^{-3}$&0&6	&X	\\
\object{RXCJ0345.7-4112}&0.06&0.67&$(2.69\pm0.67)\times10^{-7}$&0&$(4.02\pm0.54)\times10^{-3}$&0&6	&X	\\
\object{RXCJ0547.6-3152}&0.15&1.14&$(1.33\pm0.44)\times10^{-7}$&0&$(1.08\pm0.04)\times10^{-2}$&0&1,6*	&X	\\
\object{RXCJ0605.8-3518}&0.14&0.98&$(1.24\pm0.49)\times10^{-8}$&0&$(6.57\pm0.24)\times10^{-3}$&0&6	&X	\\
\object{RXCJ0616.8-4748}&0.12&0.95&$(5.83\pm1.41)\times10^{-7}$&0&$(9.95\pm0.99)\times10^{-3}$&0&6	&X	\\
\object{RXCJ0645.4-5413}&0.16&1.23&$2.29\times10^{-8}$	       &1&$(9.38\pm0.48)\times10^{-3}$&0&1,6*	&X	\\
\object{RXCJ0821.8+0112}&0.08&0.74&$(2.67\pm1.64)\times10^{-6}$&0&$(7.97\pm4.52)\times10^{-2}$&0&6	&X	\\
\object{RXCJ0958.3-1103}&0.17&1.06&$(1.67\pm3.68)\times10^{-8}$&0&$(2.97\pm0.32)\times10^{-3}$&0&1,6*	&X	\\
\object{RXCJ1044.5-0704}&0.13&0.83&$2.14\times10^{-9}$	       &1&$(3.73\pm0.20)\times10^{-3}$&0&6	&X	\\
\object{RXCJ1141.4-1216}&0.12&0.82&$(1.23\pm1.44)\times10^{-8}$&0&$(2.82\pm0.41)\times10^{-3}$&0&6	&X	\\
\object{RXCJ1236.7-3354}&0.08&0.75&$4.07\times10^{-8}$	       &1&$(3.12\pm0.42)\times10^{-3}$&0&6	&X	\\
\object{RXCJ1302.8-0230}&0.08&0.79&$(1.33\pm0.42)\times10^{-7}$&0&$(2.21\pm0.07)\times10^{-2}$&0&6	&X	\\
\object{A1689}		&0.18&1.40&$(3.38\pm2.29)\times10^{-9}$&0&$(2.03\pm0.19)\times10^{-3}$&0&1,3,6*&X	\\
\object{RXCJ1516.3+0005}&0.12&0.98&$(1.30\pm1.42)\times10^{-8}$&0&$(9.00\pm0.42)\times10^{-3}$&0&6	&X	\\
\object{RXCJ1516.5-0056}&0.12&0.86&$(9.77\pm1.89)\times10^{-7}$&0&$(1.62\pm0.10)\times10^{-2}$&0&6	&X	\\
\object{RXCJ2014.8-2430}&0.15&1.00&$(1.72\pm0.64)\times10^{-8}$&0&$(6.46\pm0.22)\times10^{-3}$&0&6	&X	\\
\object{RXCJ2023.0-2056}&0.06&0.76&$(7.82\pm6.79)\times10^{-8}$&0&$(2.14\pm0.14)\times10^{-2}$&0&6	&X	\\
\object{RXCJ2048.1-1750}&0.15&0.99&$(3.83\pm0.79)\times10^{-7}$&0&$(1.58\pm0.07)\times10^{-2}$&0&6	&X	\\
\object{RXCJ2129.8-5048}&0.08&0.91&$(1.10\pm0.69)\times10^{-7}$&0&$(5.18\pm0.10)\times10^{-2}$&0&6	&X	\\
\object{RXCJ2149.1-3041}&0.12&0.82&$(1.33\pm0.31)\times10^{-7}$&0&$(4.55\pm0.55)\times10^{-3}$&0&6	&X	\\
\object{RXCJ2217.7-3543}&0.15&1.01&$(7.94\pm2.61)\times10^{-8}$&0&$(4.36\pm0.34)\times10^{-3}$&0&6	&X	\\
\object{RXCJ2218.6-3853}&0.14&1.13&$(8.61\pm2.23)\times10^{-8}$&0&$(1.72\pm0.06)\times10^{-2}$&0&1,6*	&X	\\
\object{RXCJ2234.5-3744}&0.15&1.32&$(4.06\pm3.77)\times10^{-9}$&0&$(2.10\pm0.04)\times10^{-2}$&0&1,6*	&X	\\
\object{RXCJ2319.6-7313}&0.10&0.66&$2.29\times10^{-8}$	       &1&$(1.91\pm0.11)\times10^{-2}$&0&6	&X	\\
\object{RXCJ2157.4-0747}&0.06&0.72&$(9.67\pm2.36)\times10^{-6}$&0&$(2.15\pm0.88)\times10^{-1}$&0&6	&X	\\												
\hline \multicolumn{9}{|c|}{} \\[-1.5ex] 
\multicolumn{9}{|c|}{high-z 400SD sample} \\[0.5ex]
\hline\\[-1.2ex]
\object{1212+2733}&0.35	&1.08	&$(3.59\pm0.94)\times10^{-6}$&0&$(9.49\pm1.79)\times10^{-3}$&0&7&	C	\\
\object{0350-3801}&0.36	&0.61	&$(1.08\pm1.51)\times10^{-6}$&0&$(5.23\pm3.13)\times10^{-3}$&0&7&	C	\\
\object{0318-0302}&0.37	&0.81	&$4.84\times10^{-7}$	     &1&$(4.10\pm0.41)\times10^{-2}$&0&7&	C	\\
\object{0159+0030}&0.39	&0.82	&$7.45\times10^{-7}$	     &1&$(4.35\pm2.34)\times10^{-3}$&0&7&	C	\\
\object{0958+4702}&0.39	&0.74	&$9.22\times10^{-7}$	     &1&$(8.90\pm4.59)\times10^{-3}$&0&7&	C	\\
\object{1003+3253}&0.42	&0.93	&$(2.36\pm1.56)\times10^{-6}$&0&$(2.67\pm0.58)\times10^{-2}$&0&7&	C	\\
\object{0141-3034}&0.44	&0.54	&$2.97\times10^{-6}$	     &1&$(2.18\pm5.28)\times10^{-3}$&0&7&	C	\\
\object{1701+6414}&0.45	&0.80	&$1.63\times10^{-7}$	     &1&$(1.35\pm0.27)\times10^{-2}$&0&7&	C	\\
\object{1641+4001}&0.46	&0.68	&$(6.65\pm10.2)\times10^{-7}$&0&$(4.65\pm4.25)\times10^{-3}$&0&7&	C	\\
\object{1222+2709}&0.47	&0.73	&$1.14\times10^{-6}$	     &1&$2.51\times10^{-3}$		    &1&7&	C	\\
\object{0355-3741}&0.47	&0.82	&$(1.37\pm1.33)\times10^{-6}$&0&$(5.27\pm4.15)\times10^{-3}$&0&7&	C	\\
\object{0030+2618}&0.50	&0.66	&$1.10\times10^{-6}$	     &1&$(4.25\pm5.31)\times10^{-3}$&0&7&	C	\\
\object{1002+6858}&0.50	&0.75	&$(5.40\pm10.4)\times10^{-7}$&0&$(2.63\pm0.47)\times10^{-2}$&0&7&	C	\\
\object{1524+0957}&0.52	&0.76	&$3.11\times10^{-7}$	     &1&$(1.46\pm0.25)\times10^{-2}$&0&7&	C	\\
\object{1120+2326}&0.56	&0.67	&$5.98\times10^{-7}$	     &1&$(3.10\pm0.48)\times10^{-2}$&0&7&	C	\\
\object{1120+4318}&0.60	&0.79	&$(1.10\pm0.71)\times10^{-6}$&0&$(2.63\pm2.70)\times10^{-3}$&0&7&	C	\\
\object{1202+5751}&0.68	&0.68	&$2.08\times10^{-6}$	     &1&$(4.36\pm0.65)\times10^{-2}$&0&7&	C	\\
\object{0405-4100}&0.69	&0.66	&$(6.20\pm9.07)\times10^{-7}$&0&$(3.06\pm2.46)\times10^{-3}$&0&7&	C	\\
\object{1221+4918}&0.70	&0.88	&$(1.55\pm3.50)\times10^{-7}$&0&$(1.23\pm0.29)\times10^{-2}$&0&7&	C	\\
\object{0230+1836}&0.80	&0.75	&$1.05\times10^{-6}$	     &1&$(1.92\pm0.50)\times10^{-2}$&0&7&	C	\\
\object{0809+2811}&0.40	&0.81	&$1.03\times10^{-6}$	     &1&$(1.19\pm0.46)\times10^{-2}$&0&7&	C	\\
\object{0333-2456}&0.48	&0.66	&$(1.61\pm2.30)\times10^{-6}$&0&$4.27\times10^{-3}$		    &1&7&	C	\\
\object{1334+5031}&0.62	&0.72	&$3.31\times10^{-6}$	     &1&$(3.28\pm4.75)\times10^{-3}$&0&7&	C	\\
\object{0542-4100}&0.64	&0.81	&$(5.48\pm9.29)\times10^{-7}$&0&$(5.04\pm0.59)\times10^{-2}$&0&7&	C	\\
\object{0152-1358}&0.83	&0.72	&$(5.76\pm0.95)\times10^{-5}$&0&$(6.64\pm0.57)\times10^{-2}$&0&7&	C	\\
\object{0302-0423}&0.35	&0.90	&$7.48\times10^{-8}$	     &1&$(1.56\pm0.87)\times10^{-3}$&0&7&	C	\\
\object{1312+3900}&0.40	&0.75	&$2.27\times10^{-6}$	     &1&$(3.42\pm0.85)\times10^{-2}$&0&7&	C	\\
\object{1416+4446}&0.40	&0.70	&$3.54\times10^{-7}$	     &1&$(1.32\pm0.28)\times10^{-2}$&0&7&	C	\\
\object{0328-2140}&0.59	&0.81	&$2.25\times10^{-7}$         &1&$(5.99\pm1.97)\times10^{-3}$&0&7&	C	\\
\object{0522-3624}&0.47	&0.70	&$(1.34\pm1.46)\times10^{-6}$&0&$(3.16\pm4.01)\times10^{-3}$&0&7&	C	\\
\object{0853+5759}&0.48	&0.69	&$9.04\times10^{-7}$	     &1&$(5.24\pm0.96)\times10^{-2}$&0&7&	C	\\
\object{0926+1242}&0.49	&0.82	&$(3.12\pm5.24)\times10^{-7}$&0&$2.40\times10^{-3}$		    &1&7&	C	\\
\object{0956+4107}&0.59	&0.74	&$(1.26\pm1.01)\times10^{-6}$&0&$(7.14\pm0.57)\times10^{-2}$&0&7&	C	\\
\object{1226+3332}&0.89	&1.05	&$1.93\times10^{-7}$	     &1&$(1.11\pm0.17)\times10^{-2}$&0&7&	C	\\
\object{1354-0221}&0.55	&0.70	&$2.26\times10^{-6}$	     &1&$(1.42\pm0.55)\times10^{-2}$&0&7&	C	\\
\object{1357+6232}&0.53	&0.79	&$5.91\times10^{-7}$	     &1&$(1.25\pm0.29)\times10^{-2}$&0&7&	C	\\		
\hline \multicolumn{9}{|c|}{} \\[-1.5ex] 
\multicolumn{9}{|c|}{high-z SPT sample} \\[0.5ex]
\hline\\[-1.2ex]
\object{SPT-CLJ0000-5748}&0.74&	1.00&$3.55\times10^{-7}$	  &1&$(8.34\pm3.25)\times10^{-3}$&0&8	&C	\\
\object{SPT-CLJ0509-5342}&0.46&	1.04&$4.27\times10^{-7}$	  &1&$(3.13\pm1.33)\times10^{-3}$&0&8	&C	\\
\object{SPT-CLJ0516-5430}&0.29&	1.12&$(7.91\pm3.33)\times10^{-7}$ &0&$(2.88\pm0.22)\times10^{-2}$&0&1*,2,8&X	\\
\object{SPT-CLJ0528-5300}&0.76&	0.74&$(2.92\pm2.37)\times10^{-6}$ &0&$(7.35\pm3.69)\times10^{-3}$&0&8	&C	\\
\object{SPT-CLJ0533-5005}&0.88&	0.59&$(4.36\pm12.70)\times10^{-6}$&0&$(1.42\pm0.93)\times10^{-2}$&0&8	&C	\\
\object{SPT-CLJ0546-5345}&1.07&	0.76&$7.08\times10^{-7}$	  &1&$(6.75\pm2.25)\times10^{-3}$&0&8	&C	\\
\object{SPT-CLJ0551-5709}&0.42&	0.79&$1.17\times10^{-6}$	  &1&$(2.75\pm0.60)\times10^{-2}$&0&8	&C	\\
\object{SPT-CLJ0559-5249}&0.61&	1.01&$(1.06\pm1.05)\times10^{-6}$ &0&$(2.93\pm0.41)\times10^{-2}$&0&8	&X	\\
\object{SPT-CLJ2331-5051}&0.57&	0.89&$(2.47\pm3.26)\times10^{-7}$ &0&$(6.18\pm2.10)\times10^{-3}$&0&8	&C	\\
\object{SPT-CLJ2337-5942}&0.78&	0.99&$1.55\times10^{-7}$	  &1&$(4.32\pm4.34)\times10^{-3}$&0&8	&C	\\
\object{SPT-CLJ2341-5119}&1.00&	0.82&$3.89\times10^{-7}$	  &1&$(5.78\pm1.87)\times10^{-3}$&0&8	&C	\\
\object{SPT-CLJ2342-5411}&1.08&	0.60&$8.71\times10^{-7}$	  &1&$2.45\times10^{-3}$		    &1&8	&C	\\
\object{SPT-CLJ2332-5358}&0.32&	1.17&$(6.23\pm16.90)\times10^{-7}$&0&$(4.38\pm3.42)\times10^{-3}$&0&8	&X	\\
\object{SPT-CLJ2355-5056}&0.35&	0.97&$(2.20\pm4.02)\times10^{-7}$ &0&$(6.42\pm2.11)\times10^{-3}$&0&8	&C	\\
\object{SPT-CLJ2359-5009}&0.76&	0.83&$1.32\times10^{-6}$	  &1&$(1.44\pm0.58)\times10^{-2}$&0&8	&C	\\
\label{SampleInfo}			    				   
\end{longtable} 
\tablefoot{Column 2: cluster redshift; Column 3: \r500 in Mpc estimated from the formula given by \citet{Arnaud2005}; Column 4/6: bias- and background-corrected \P3P0 and $w$ values calculated in an aperture of \r500 including the central region. Errors are 1-$\sigma$ uncertainties; Column 5/7: flags for upper limits where 0 indicates a significant detection and 1 an upper limit. For flag 1 \P3P0/$w$ and its error are the same. The upper limit is calculated as described in Sect. \ref{Sect3}; Column 8: Reference. In case of multiple references, * indicates the temperature source for the \r500 calculation; Column 9: Flag indicating whether an XMM-Newton (X) or CHANDRA (C) image was used in the analysis. }
\tablebib{(1) LoCuSS: \citet{Zhang2008}; (2) REFLEX-DXL: \citet{Zhang2006}; (3) \citet{Snowden2008}; (4) \citet{Arnaud2005};\\ (5) \citet{Buote1996}; (6) REXCESS: \citet{Boehringer2009}; (7) high-z 400SD sample: \citet{Vikhlinin2009}; (8) high-z SPT sample: \citet{Andersson2011}.}
\end{center}

\end{document}